\documentclass{JHEP3}

\usepackage{amsmath,amstext,amssymb}
\usepackage{psfrag,graphicx,subfigure}
\usepackage{cite}

\def\caln{{\cal N}}

\def\calo{{\cal O}}

\def\calj{{\cal J}}
\def\ie{{i.\,e.\ }}
\def\eg{{e.\,g.\ }}
\def\dt{\tilde d}
\def\mut{\tilde\mu}
\def\ft{\tilde f}
\def\Nt{\tilde N}
\def\vrho{\varrho}
\def\dt{\tilde d}
\def\At{\tilde A}
\def\pt{\tilde p}
\def\Xt{\tilde X}
\def\rt{\tilde r}
\def\Lt{\tilde L}
\def\mt{\tilde m}
\def\phit{\tilde \phi}
\def\psit{\tilde \psi}
\def\mut{\tilde \mu}
\def\wt{\tilde w}
\def\It{\tilde I}

\def\cala{{\cal A}}
\def\calw{{\cal W}}
\def\calf{{\cal F}}

\def\calr{{\cal R}}

\def\del{\partial}

\def\str{\mathrm{Str}}
\def\ee{{\mathrm e}}
\def\ii{{\mathrm i}}

\newcommand{\dd}{\mathrm{d}}

\newcommand{\eins}{\mbox{$1 \hspace{-1.0mm} {\bf l}$}}

\title{Holographic  Superfluidity in Imbalanced Mixtures}

\author{Johanna Erdmenger, Viviane Grass, Patrick Kerner and Thanh Hai Ngo \\ Max-Planck-Institut f\"ur Physik (Werner-Heisenberg-Institut)\\
  F\"ohringer Ring 6, 80805 M\"unchen, Germany\\ \email{jke, grass, pkerner, ngo@mppmu.mpg.de}}

\abstract{We construct superfluid black hole solutions with two
chemical potentials.  By
analogy with QCD, the two chemical potentials correspond to the baryon
and isospin symmetries, respectively. We consider two systems: 
the back-reacted $U(2)$
Einstein-Yang-Mills theory in 4+1 dimensions and the 9+1-dimensional
D$3$/D$7$ brane setup with two coincident D$7$-brane probes. 
In the D$7$-brane model, the
identification of baryon and isospin chemical potential 
is explicit since the dual field theory is explicitly
known. Studying the phase diagram, we find in both systems a quantum
phase transition at a critical ratio of the two chemical
potentials. However the quantum phase transition is different in the
two systems: In the D$3$/D$7$ brane setup we always find a second
order phase transition, while in the Einstein-Yang-Mills theory,
depending on the strength of the back-reaction, we
obtain a continuous or first order transition. We expect the
continuous quantum phase transition to be BKT-like. We comment on the 
origin of this
differing behavior in these apparently very similar models and 
compare to phenomenological systems.}

\date{\today}


\preprint{MPI-2011-19}

\begin{document}

%%%%%%%%%%%%%%%%%%%%%%%%%%%%%%%% I N T R O
% !TEX root = SFisobar.tex
%-------------------------------------------------------------------------------------------------------------------------------------------
\section{Introduction}
\label{sec:Introduction}
%-------------------------------------------------------------------------------------------------------------------------------------------

In condensed matter physics, phenomenologically interesting systems are often strongly coupled. Famous examples are high $T_c$-superconductors and ultra-cold Fermi gases. Strong\-ly correlated systems may undergo a phase transition at zero temperature which is driven by quantum fluctuations and thus named quantum phase transition \cite{Sachdevbook}. Especially interesting are continuous quantum phase transitions which feature a quantum critical point. This quantum critical point influences the phase diagram also at non-zero temperature. In this influenced region, the quantum critical region, the system may be described by a critical theory even at finite temperature \cite{Sachdev:2009ap,Sachdev:2010ch}.

Gauge/gravity duality \cite{Maldacena:1997re, Aharony:1999ti} provides a novel method for studying strongly correlated systems at finite temperature and densities. As such it should help to understand the systems described above \cite{Hartnoll:2009sz, Herzog:2009xv, McGreevy:2009xe}. Indeed remarkable progress was made in the application of gauge/gravity duality towards the description of superfluids and superconductors following the results of \cite{Hartnoll:2008vx},  as well as of (non-) Fermi liquids \cite{Liu:2009dm}.

So far systems showing the transition to a holographic superfluid have only been  considered with one control parameter, usually the ratio of chemical potential and temperature\footnote{An exception is \cite{Faulkner:2010gj} where a second control parameter induced by multi-trace deformations is considered.}. Naturally in such systems the phase transition is at a finite temperature and thus these systems have no quantum phase transition. In this paper we study a holographic system which shows a transition to a superfluid phase and features an additional chemical potential as a second control parameter. Since our starting point is a conformal field theory, we will construct two dimensionless control parameters: the ratio of one chemical potential to temperature and the ratio of the two chemical potentials. In such systems the ratio of temperature to one chemical potential at which the phase transition occurs may be tuned to zero by the second control parameter. Thus these systems may contain quantum phase transitions.

\begin{figure}
\centering
\subfigure[]{\includegraphics[width=0.45\textwidth]{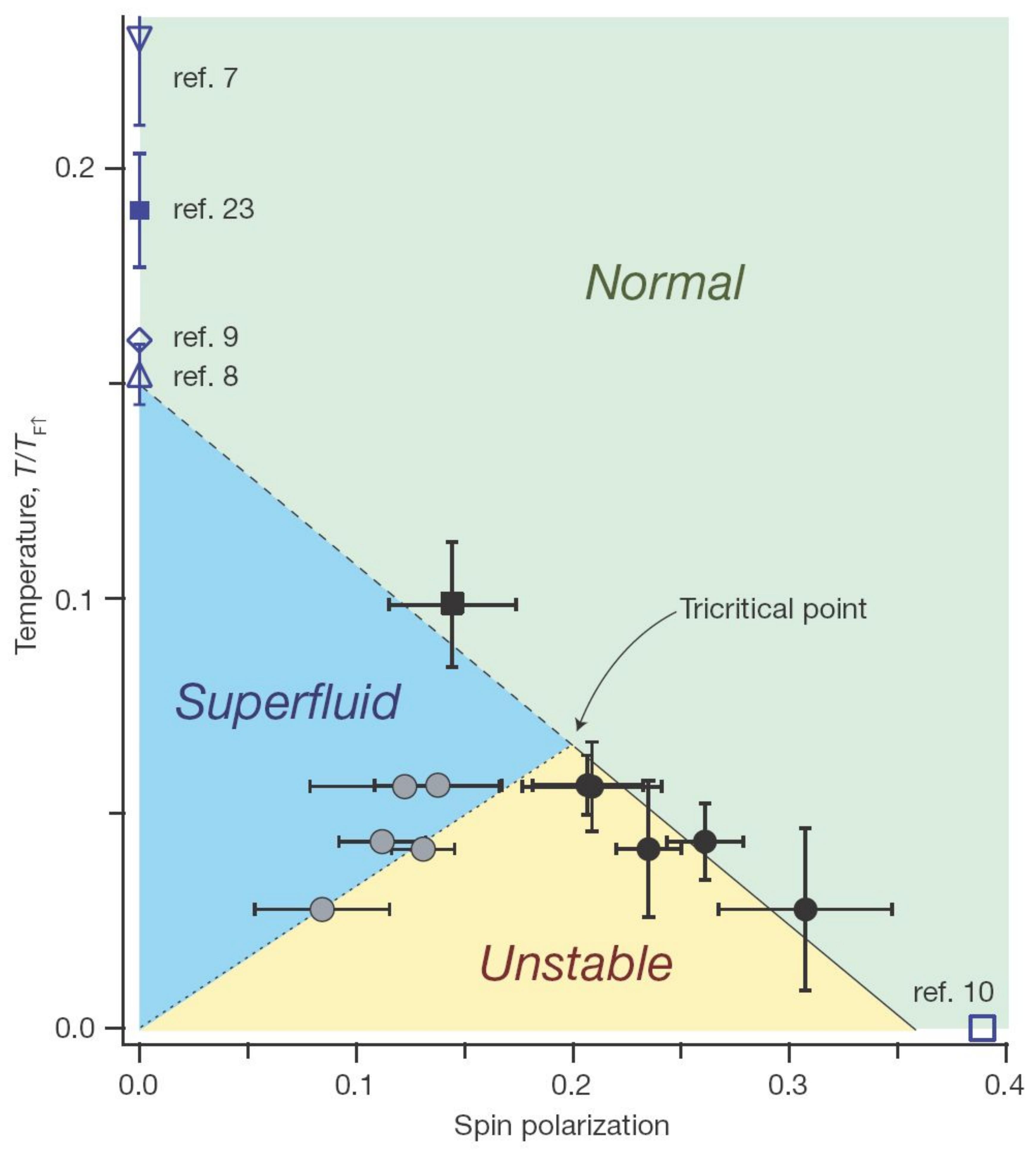}}\hfill
\subfigure[]{\includegraphics[width=0.45\textwidth]{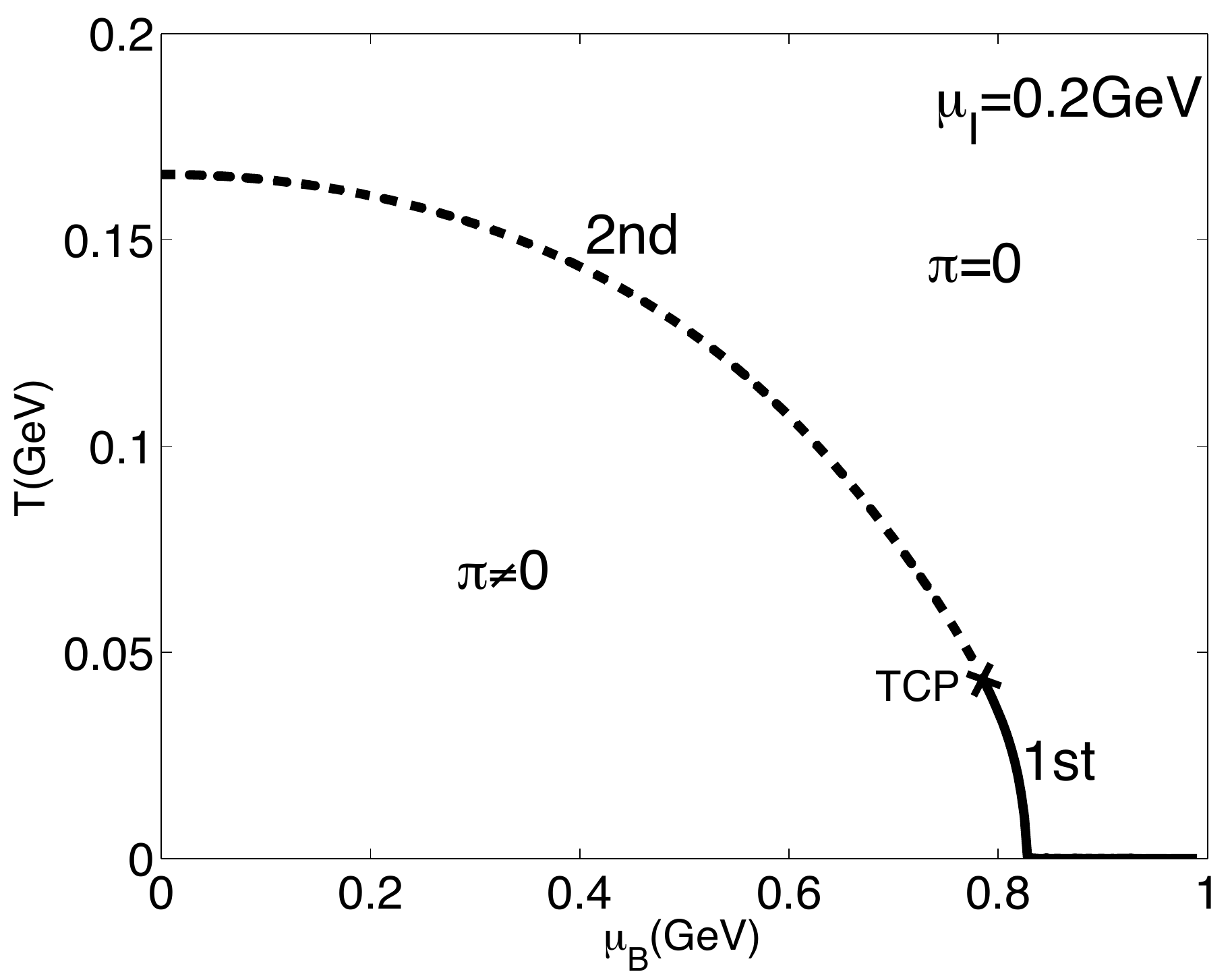}}
\caption{Phase diagrams of real world systems: (a) Imbalanced Fermi mixture in the canonical ensemble \cite{Shin:2007pd}:  The spin polarization is the thermodynamic conjugated variable to the ratio of the chemical potentials favoring the different spins. (b) QCD at finite baryon and isospin chemical potential \cite{He:2005nk}. In both phase diagrams we observe a superfluid phase at small temperature and small ratio of two chemical potentials. In addition in both diagrams the phase transition is second order for large temperature and becomes first order at low temperatures. Both diagrams show a first order quantum phase transition. Both figures are reproduced by kind permission of the authors.}
\label{fig:phasediagrampheno}
\end{figure}

Systems in which two chemical potentials can be tuned are often called imbalanced mixtures since two kinds of particles are present in imbalanced numbers. Examples are imbalanced Fermi mixtures where fermions with spin up and spin down are imbalanced \cite{Shin:2007pd},  and QCD at finite baryon and isospin chemical potential where for instance up and down quarks are imbalanced \cite{He:2005nk} (see also \cite{Chernodub:2011fr}). Interestingly the phase diagrams of both these systems are very similar (see figure \ref{fig:phasediagrampheno}). In both systems there is a superfluid state at low temperatures and at certain ratios of the two chemical potentials. In addition also the order of the phase transition agrees in both examples: At low temperatures (also at zero temperature) the transition is first order while at higher temperatures the transition becomes second order. Is it possible that there is an universal structure which relates these two different systems? 

\medskip

In this paper we holographically study field theories which are expected to be similar to theories which describe the systems discussed above. Our field theories have a global $U(2)$ symmetry which may be split into $U(1)\times SU(2)$. This allows us to switch on two chemical potentials: one for the overall $U(1)$ and one for a diagonal $U(1)$ inside $SU(2)$. In analogy to QCD, the chemical potential for the overall $U(1)$ is the baryon chemical potential, while the one for the diagonal $U(1)$ is identified with the isospin chemical potential. On the gravity side, we realize the $U(2)$ gauge theory in two different ways: As a first model, we consider the $U(2)$ Einstein-Yang-Mills (EYM) theory. In this model we allow the gauge fields to back-react on the geometry in order to get a coupling between the overall $U(1)$ gauge fields and the $SU(2)$ gauge fields. As a second model, we consider  the D$3$/D$7$ brane setup with two coincident D$7$-brane probes which feature the $U(2)$ gauge theory. In this model the interaction between the overall $U(1)$ and the $SU(2)$ gauge fields is obtained by the Dirac-Born-Infeld action.

So far, holographic superfluidity in the Einstein-Yang-Mills theory in the presence of just an isospin chemical potential has been studied in \eg \cite{Gubser:2008zu,Gubser:2008wv,Basu:2009vv,Ammon:2009xh}. In the probe approximation \cite{Gubser:2008wv}, \ie the gauge fields do not influence the metric, a second order phase transition to a state is found which spontaneously breaks an Abelian symmetry. This spontaneous breaking creates a superfluid. In \cite{Ammon:2009xh}, the back-reaction of the gauge field on the metric has been added to this scenario. By increasing the back-reaction, the critical temperature decreases. Beyond a critical strength of the back-reaction, the phase transition is first order. There is a maximal value for the back-reaction beyond which the transition to the superfluid phase is not possible. 

The simple bulk action of the Einstein-Yang-Mills theory has the great virtue of being universal: The results may be true for many different dual field theories independently of their dynamics. Unfortunately this simple construction does not allow to identify the dual field theory explicitly. However it has been shown in \cite{Ammon:2008fc,Basu:2008bh,Ammon:2009fe} that the Einstein-Yang-Mills system can be embedded into string theory by considering the D$3$/D$7$ brane setup (see \eg \cite{Karch:2002sh,Erdmenger:2007cm}). The dual field theory of the D$3$/D$7$ brane setup is known explicitly: It is  $\caln=4$ Super-Yang-Mills theory coupled to $\caln=2$ hypermultiplets. In this setup we work in the probe approximation, \ie we consider $N_c\gg 1$ D$3$-branes which generate the background metric $AdS_5\times S^5$ and embed $N_f=2$ D$7$-branes into the background space. The embedding of the D$7$-branes generates degrees of freedom which transform in the fundamental representation of the gauge group, the $\caln=2$ hypermultiplets, which we denote as quarks in analogy to QCD. 
Here we have two quark flavors. Since the dual field theory is known explicitly, also our identification of the two chemical potentials as corresponding to the $U(1)$ baryon and $SU(2)$ isospin symmetries is explicitly realized. 
In this theory, mesonic bound states of the fundamental degrees of freedom are formed. The transition to the superfluid state is related to the condensation of vector mesons which spontaneously break an Abelian symmetry \cite{Ammon:2008fc}.

In both systems, \ie in the EYM and in the D$3$/D$7$ model, the mechanism of breaking the Abelian symmetry is the same in the bulk: A non-zero vev of the time component of the gauge field $A_t$ induces a chemical potential on the boundary theory. By fixing a gauge, we can choose the $SU(2)$ gauge field in the direction of the third Pauli matrix to be non-zero, \ie $A_t^3\not=0$. This breaks the $SU(2)$ symmetry down to an Abelian symmetry which we call $U(1)_3$. Beyond a critical value of the chemical potential, the systems become unstable against fluctuations of the gauge field pointing in some other direction inside the $SU(2)$, for instance $A_x^1$ (see \eg \cite{Gubser:2008wv} for the Einstein-Yang-Mills system and \cite{Erdmenger:2008yj} for the D$3$/D$7$ brane setup). This instability is cured by the condensation of this gauge field $A_x^1$ which then breaks the $U(1)_3$ symmetry. In the boundary theory the non-trivial profile of the gauge field $A_x^1$ induces a vev of the current $\langle J_x^1\rangle$, but no source. Thus the breaking of the $U(1)_3$ symmetry is spontaneous and the order parameter for the transition to the superfluid phase is given by $\langle J_x^1\rangle$. For the D$3$/D$7$ brane setup we can explicitly write down the field content of the order parameter \cite{Ammon:2008fc},
\begin{equation}
  \label{eq:fieldorderparameter}
    J^1_x\propto \bar\psi\sigma^1\gamma_x\psi+\phi\sigma^1\del_x\phi=\bar\psi_u\gamma_x \psi_d+\bar \psi_d\gamma_x \psi_u+\text{bosons} \,,
 \end{equation}
where $\psi=(\psi_u,\psi_d)$ and $\phi=(\phi_u,\phi_d)$ are the quarks and squarks duplet, respectively, $\sigma^i$ denote the Pauli matrices and $\gamma_\mu$ the Dirac matrices.
\begin{figure}
\centering
\subfigure[]{\includegraphics[width=0.33\textwidth]{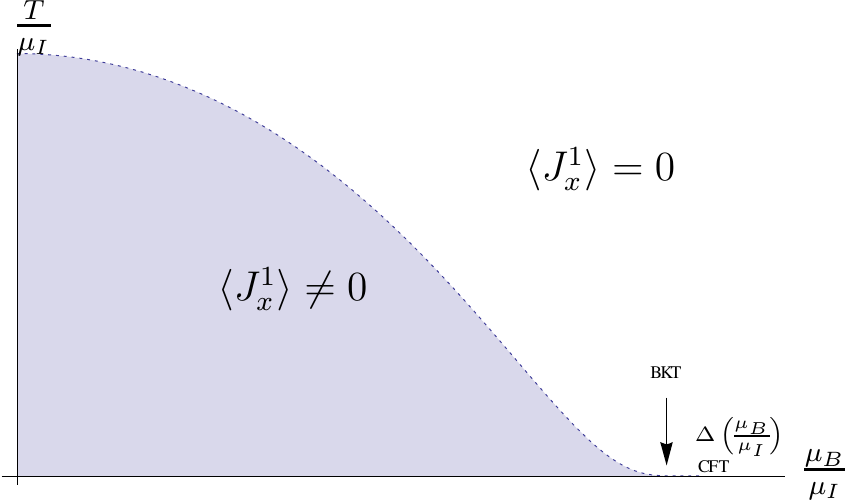}}\hfill
\subfigure[]{\includegraphics[width=0.33\textwidth]{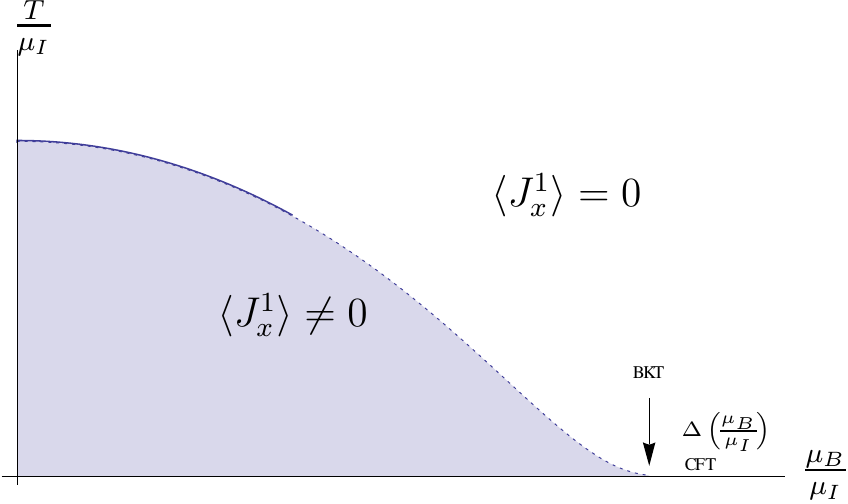}}\hfill
\subfigure[]{\includegraphics[width=0.33\textwidth]{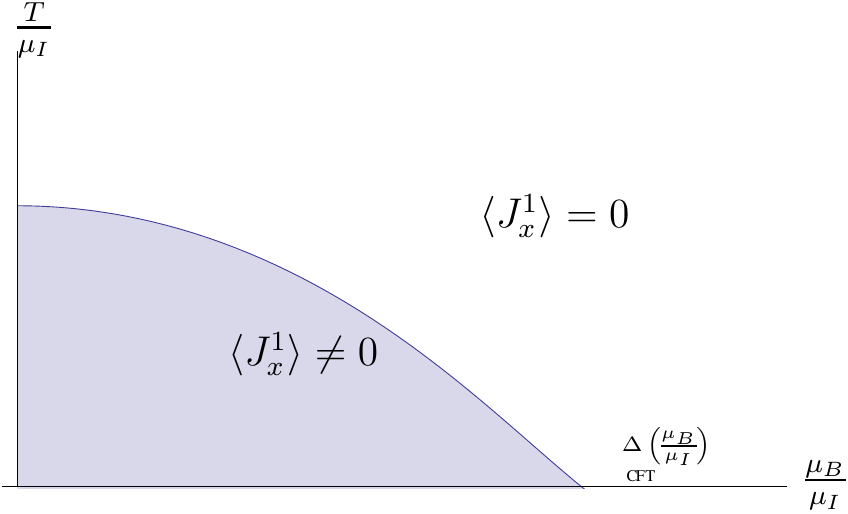}}
\caption{Sketch of the phase diagrams for the Einstein-Yang-Mills system for different strength of the back-reaction: In the white region the system is in the normal phase while in the blue region it is in the superfluid phase. The solid line marks a first oder phase transition and the dotted line a second order phase transition. In the normal phase at zero temperature the dual field theory contains an emergent one-dimensional CFT in the IR and the IR dimension of the operator depends on the ratio of the chemical potentials. For small back-reaction (a), the phase transition is second order for finite temperatures and we expect the quantum phase transition to be BKT-like. For intermediate back-reaction (b), there is a first order phase transition at large temperatures. At low temperatures the behavior is as for small back-reaction. For large back-reaction (c), the phase transition is always first order. Also the quantum phase transition is first order.}
\label{fig:sketchEYM}
\end{figure}

\begin{figure}
\centering
\includegraphics[width=0.75\textwidth]{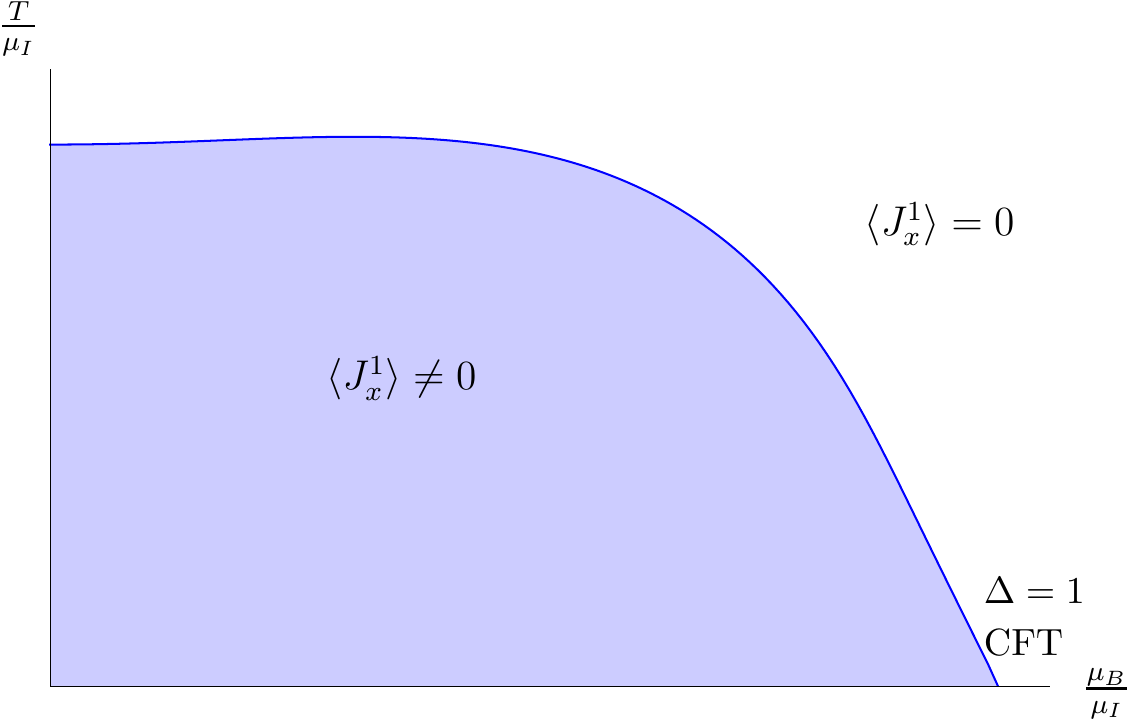}
\caption{Sketch of the phase diagram for the D$3$/D$7$ brane setup: In the white region the system is in the normal phase while in the blue region it is in the superfluid phase. The critical temperature is first increasing and later monotonically decreasing as $\mu_B/\mu_I$ is increased. The phase transition is always second order. We also expect the quantum phase transition to be second order with mean-field exponents. There is also an emergent CFT in the IR at zero temperature in the normal phase. However, the IR dimension of the dual operator does not depend on $\mu_B/\mu_I$ and remains constant equal to $1$.}
\label{fig:sketchD3D7}
\end{figure}

In this paper, in addition to the time component of the gauge field inside the $SU(2)$ $A_t^3$, we switch on the time component of the $U(1)$ gauge field $\cala_t$ which induces the second control parameter, the baryon chemical potential. Tuning the two control parameters we can map out the phase diagram of both systems and find interesting similarities and differences (see figures \ref{fig:sketchEYM} and \ref{fig:sketchD3D7} for a sketch of the phase diagrams). In both cases the critical temperature where the phase transition occurs is finite at zero baryon chemical potential. By increasing the baryon chemical potential, we can tune the critical temperature to zero and we obtain a quantum phase transition. However it is interesting that the details of the phase diagram are very different for the two systems, although they are expected to be dual to very similar field theories. For instance, the local as well as global symmetries match. The differences in the phase diagram are: In the Einstein-Yang-Mills theory (see figure~\ref{fig:sketchEYM}) the critical temperature is monotonically decreasing as we increase the baryon chemical potential, while in the D$3$/D$7$ brane setup (see figure~\ref{fig:sketchD3D7}) the critical temperature first increases as the baryon chemical potential is increased. In addition in the Einstein-Yang-Mills setup, the system exhibits first and second order phase transitions depending on the strength of the back-reaction, while in the D$3$/D$7$ brane setup we obtain only second order phase transitions. Thus the question arises: What is the crucial difference between the systems which induces the different phase transitions?

From the construction there is one  obvious difference. In the Einstein-Yang-Mills system, the $U(1)$ and $SU(2)$ gauge fields only couple indirectly via the metric. In the field theory this means that the coupling of the currents which are dual to the gauge fields only occurs due to gluon loops. In the D$3$/D$7$ brane setup these loops are neglected due to the probe approximation. In this case the field theory currents directly interact with each other. These interactions are induced by the non-linear terms of the DBI action. Due to this difference it is understandable that the phase transitions may be different. The different couplings of the gauge fields to each other may lead to different RG flows and therefore to different IR physics which lead to differences in the phase diagram.

In addition we find an interesting difference in the  origin of the quantum critical point in the systems. In the Einstein-Yang-Mills setup we can pinpoint the origin of the instability to the violation of the Breitenlohner-Freedman bound in an IR $AdS_2$ region. This $AdS_2$ region shows up as the near horizon region of the extremal Reissner-Nordstr\"om black hole. According to the AdS/CFT dictionary, the dual field theory thus contains a one-dimensional CFT in the IR (see figure~\ref{fig:sketchEYM}). It is also important that the IR dimension of the dual operator depends on the ratio of the chemical potentials, such that the dimension can be tuned to an unstable value. In \cite{Kaplan:2009kr} it is argued that the violation of the Breitenlohner-Freedman bound will lead to a BKT-like phase transition. A common feature for this kind of transition seems to be the turning point in the phase diagram, such that the critical temperature slowly goes to zero as the ratio of baryon to isospin chemical potential is increased. In contrast to this behavior, the critical temperature in the D$3$/D$7$ brane setup goes to zero linearly. In this second model we do not obtain a violation of the Breitenlohner-Freedman bound in the IR, since the IR dimension of the dual operator does not depend on the ratio of the chemical potentials. Therefore we expect that the quantum phase transition is second order with mean field exponents. 

A similar difference occurs for the phase transition of chiral symmetry breaking via magnetic catalysis \cite{Filev:2007gb,Erdmenger:2007bn}, if we compare this transition for the D$3$/D$5$ brane setup to the same transition for the D$3$/D$7$ brane setup \cite{Evans:2010hi, Evans:2010iy}. In the D$3$/D$5$ model the quantum phase transition is BKT-like, while in the D$3$/D$7$ model it is second order. Here the difference between the two systems is more obvious: In the D$3$/D$5$ model both control parameters, the magnetic field and the baryon density, have mass dimension two. In the D$3$/D$7$ model the magnetic field still has mass dimension two but the baryon density has now mass dimension three such that the dimensions of the control parameters do not match. It is expected that the BKT-like transition only occurs if the two control parameters have the same dimension (see \cite{Jensen:2010ga,Pal:2010gj,Jensen:2010vx,Evans:2010np}). This is in contrast to field theories with two chemical potentials where the dimensions of the two control parameters match independently of the spacetime dimensions. This is a great advantage since these systems always satisfy the necessary condition for a BKT-like transition. However, this condition cannot be sufficient as our result in the D$3$/D$7$ brane setup shows. 

Comparing the phase diagrams obtained in our models (see figure~\ref{fig:sketchEYM} and \ref{fig:sketchD3D7}) with the one obtained in imbalanced Fermi mixtures \cite{Shin:2007pd} and QCD at finite baryon and isospin chemical potential \cite{He:2005nk} (see figure \ref{fig:phasediagrampheno}), we see some similarities. In all cases the critical temperature is finite if the second control parameter, in our case the baryon chemical potential, is zero. By increasing the second control parameter we can tune the critical temperature to zero and we obtain a quantum phase transition. This seems to be a universal behavior for  systems with two control parameters. However in imbalanced Fermi mixtures and QCD at finite baryon and isospin chemical potential shown in figure~\ref{fig:phasediagrampheno} the order of the phase transition is different from that in our models. In the models of figure~\ref{fig:phasediagrampheno},  the phase transition is second order at large temperatures and becomes first order at low temperatures. On the other hand, in the holographic models this is different: For large back-reaction the behavior in the Einstein-Yang-Mills system is completely opposite. The phase transition is first order at large temperatures and becomes continuous at small temperatures. Also for small back-reaction, we find a continuous quantum phase transition instead of a discontinuous one.

The difference in the order of the quantum phase transition may be related to the different behavior of the normal phase at zero temperature. For instance the BKT-like transition in the Einstein-Yang-Mills setup is possible since the theory is conformal in the IR and the IR dimension of the dual operator depends on $\mu_B/\mu_I$. In \cite{Kaplan:2009kr} a BKT-like transition has been discussed in conformal field theories. The transition may occur if two fixed points of the $\beta$-function annihilate. The Einstein-Yang-Mills setup is the only one considered here which is conformal in the IR with tunable IR dimension of the dual operator, unlike both the models of figure~\ref{fig:phasediagrampheno} and the D$3$/D$7$ setup, such that a different order of the phase transition is plausible. In the D$3$/D$7$ probe brane setup we do not observe any change in the order of the phase transition which is always second order. Therefore, by comparing the different models, we conclude that the order of the phase transition is not universal and depends on the precise form of the interaction.

\medskip

The paper is arranged in the following way: In section \ref{sec:Einstein-Yang-Mills-Theory} we study the back-reacted Einstein-Yang-Mills theory. In section \ref{sec:Holographic-Setup} we present its action and equations of motion. In the normal phase the solution is given by the Reissner-Nordstr\"om black hole while in the superfluid phase the equations of motion are solved numerically by the shooting method. In section \ref{sec:Thermodynamics} we determine the thermodynamic quantities and construct the phase diagram in \ref{sec:Phase-diagram}. In section \ref{sec:Zero-Temperature-Solutions-Backreact} we discuss the phase diagram at zero temperature in detail and find an analytic expression for the quantum critical point. In section \ref{sec:The-semi-probe-limit} we neglect the back-reaction of the Yang-Mills fields and analytically construct solutions in the superfluid phase for small baryon chemical potentials. This analytic solution determines the phase structure.

 In section \ref{sec:d3/d7-brane-setup} we investigate the D$3$/D$7$ brane setup. The embedding of the D$7$-branes is discussed in section \ref{sec:backgr-brane-conf}. Its action and equations of motion are given in section \ref{sec:dbi-action-equations}. In section \ref{sec:Thermodynamicsd3d7} we obtain the thermodynamic quantities and construct the phase diagram in section \ref{sec:Phase-diagramd3d7}. Zero temperature solutions and the origin of the quantum phase transition are discussed in section \ref{sec:Zero-Temperature-Solutions}. 
 
 We conclude in section~\ref{sec:Conclusion}.

%%%%%%%%%%%%%%%%%%%%%%%%%%%%%%%% T H E     M O D E L
% !TEX root = SFisobar.tex

%-------------------------------------------------------------------------------------------------------------------------------------------
\section{Einstein-Yang-Mills Theory}
\label{sec:Einstein-Yang-Mills-Theory}
%-------------------------------------------------------------------------------------------------------------------------------------------

%-------------------------------------------------------------------------------------------------------------------------------------------
\subsection{Action and equations of motion}
\label{sec:Holographic-Setup}
%-------------------------------------------------------------------------------------------------------------------------------------------
In this section we consider the $U(2)$ Einstein-Yang-Mills theory in $(4+1)$-dimensional asymptotically AdS space. The action is
\begin{equation}
\label{eq:actionmodel}
S=\int\!\dd^5 x \sqrt{-g}\left[\frac{1}{2\kappa_5^2}(\calr-\Lambda) -\frac{1}{4\hat{g}_\text{MW}^2}\calf_{\mu\nu}\calf^{\mu\nu}-\frac{1}{4\hat{g}_\text{YM}^2}F^a_{\mu\nu}F^{a\mu\nu}\right] \,,
\end{equation}
where $\kappa_5$ is the five-dimensional gravitational constant, $\Lambda=-12/R^2$ is the cosmological constant, with $R$ being the AdS radius, $\hat{g}_\text{MW}$ the Maxwell and $\hat{g}_\text{YM}$ the Yang-Mills coupling. The $U(2)$ gauge field is split into an $SU(2)$ part with field strength tensor
\begin{equation}
\label{eq:FSU(2)}
F^a_{\mu\nu}=\del_\mu A^a_\nu-\del_\nu A^a_\mu +\epsilon^{abc}A^b_\mu A^c_\nu\,,
\end{equation}
where $\epsilon^{abc}$ is the total antisymmetric tensor and $\epsilon^{123}=+1$, and into an $U(1)$ part with field strength tensor
\begin{equation}
\label{eq:FU(1)}
\calf_{\mu\nu}=\del_\mu \cala_\nu-\del_\nu\cala_\mu\,.
\end{equation}

The Einstein and Yang-Mills equations derived from the above action are
\begin{equation}
\label{eq:Einstein-Yang-Mills}
\begin{split}
\calr_{\mu\nu}+\frac{4}{R^2}g_{\mu\nu}&=\kappa_5^2\left(T_{\mu\nu}-\frac{1}{3}T^\rho_\rho g_{\mu\nu}\right)\,,\\
\nabla_\mu F^{a\mu\nu}&=-\epsilon^{abc}A^b_\mu F^{c\mu\nu}\,,\\
\nabla_\mu\calf^{\mu\nu}&=0\,,
\end{split}
\end{equation}
where the Yang-Mills energy-momentum tensor $T_{\mu\nu}$ is
\begin{equation}
\label{eq:TYangMills} 
T_{\mu\nu}=\frac{1}{\hat{g}_\text{YM}^2}\left[F^a_{\mu\rho}{F^a_\nu}^\rho-\frac{1}{4}g_{\mu\nu}F^a_{\sigma\rho}F^{a\sigma\rho}\right]
+\frac{1}{\hat{g}_\text{MW}^2}\left[\calf_{\mu\rho}{\calf_\nu}^\rho-\frac{1}{4}g_{\mu\nu}\calf_{\sigma\rho}\calf^{\sigma\rho}\right]\,.
\end{equation}
Following \cite{Ammon:2009xh}, to construct charged black hole solutions with a vector hair we choose a gauge field ansatz
\begin{equation}
\label{eq:gaugefieldansatz}
\begin{split}
A=&\phi(r)\tau^3\dd t+w(r)\tau^1\dd x\,,\\
\cala=&\psi(r)\dd t\,.
\end{split}
\end{equation}
The motivation for this ansatz is as follows: In the field theory we introduce a baryon and isospin chemical potential by the the boundary values of the time components of the gauge fields, $\phi$ and $\psi$. This breaks the $U(2)$ symmetry down to a diagonal $U(1)$ which is generated by $\tau^3$. We denote this $U(1)$ as $U(1)_3$. In order to study the transition to the superfluid state, we allow solutions with non-zero $\langle J_x^1\rangle$ such that we include the dual gauge field $A_x^1=w$ in the gauge field ansatz. Since we consider only isotropic and time-independent solutions in the field theory, the gauge fields exclusively depend on the radial coordinate $r$. With this ansatz the Yang-Mills energy-momentum tensor in \eqref{eq:TYangMills} is diagonal. Solutions with $\langle J_x^1\rangle\not=0$ also break the spatial rotational symmetry $SO(3)$ down to $SO(2)$ \footnote{Note that the finite temperature and chemical potential already break the Lorentz group down to $SO(3)$.} such that our metric ansatz will respect only $SO(2)$. Since the Yang-Mills energy-momentum tensor is diagonal, a diagonal metric is consistent. Following \cite{Manvelyan:2008sv,Ammon:2009xh}, our metric ansatz is
\begin{equation}
\label{eq:metricansatz}
\dd s^2=-N(r)\sigma(r)^2\dd t^2+\frac{1}{N(r)}\dd r^2+r^2f(r)^{-4}\dd x^2+r^2f(r)^2\left(\dd y^2+\dd z^2\right)\,,
\end{equation}
with $N(r)=-2m(r)/r^2+r^2/R^2$.

Inserting our ansatz into the Einstein and Yang-Mills equations leads to six equations of motion for $m(r), \sigma(r),f(r),\phi(r),w(r),\psi(r)$ and one constraint equation from the $rr$ component of the Einstein equations. The dynamical equations may be written as
\begin{equation}
\label{eq:eomsbr}
\begin{split}
m'&=\frac{\alpha^2_\text{YM}rf^4w^2\phi^2}{6N\sigma^2}+\frac{r^3(\alpha_\text{YM}^2 \phi'^2+\alpha_\text{MW}^2\psi'^2)}{6\sigma^2}+N\left(\frac{r^3f'^2}{f^2}+\frac{\alpha_\text{YM}^2}{6}rf^4w'^2\right)\,,\\
\sigma'&=\frac{\alpha_\text{YM}^2f^4w^2\phi^2}{3rN^2\sigma}+\sigma\left(\frac{2rf'^2}{f^2}+\frac{\alpha_\text{YM}^2f^4w'^2}{3r}\right)\,,\\
f''&=-\frac{\alpha_\text{YM}^2f^5w^2\phi^2}{3r^2N^2\sigma^2}+\frac{\alpha_\text{YM}^2f^5w'^2}{3r^2}-f'\left(\frac{3}{r}-\frac{f'}{f}+\frac{N'}{N}+\frac{\sigma'}{\sigma}\right)\,,\\
\phi''&=\frac{f^4w^2\phi}{r^2N}-\phi'\left(\frac{3}{r}-\frac{\sigma'}{\sigma}\right)\,,\\
w''&=-\frac{w\phi^2}{N^2\sigma^2}-w'\left(\frac{1}{r}+\frac{4f'}{f}+\frac{N'}{N}+\frac{\sigma'}{\sigma}\right)\,,\\
\psi''&=-\psi'\left(\frac{3}{r}-\frac{\sigma'}{\sigma}\right)\,.
\end{split}
\end{equation}
The equations of motion are invariant under five scaling transformations (invariant quantities are not shown),
\begin{equation}
\label{eq:scaltrans}
\begin{aligned}
&(I) &&\sigma\to\lambda\sigma,\quad\phi\to\lambda\phi\,,\quad\psi\to\lambda\psi \,,\\
&(II)&& f\to\lambda f,\quad w\to\lambda^{-2}w\,,\\
&(III)&&r\to\lambda r,\quad m\to\lambda^4m,\quad w\to\lambda w,\quad \phi\to\lambda\phi, \quad \psi\to\lambda\psi\,,\\
&(IV)&&r\to\lambda r,\quad m\to\lambda^2 m,\quad R\to\lambda R,\quad\phi\to\lambda^{-1}\phi\,,\quad\psi\to\lambda^{-1}\psi\,,\\
& &&\alpha_\text{YM}\to\lambda\alpha_\text{YM}\,,\quad \alpha_\text{MW}\to\lambda\alpha_\text{MW}\,,\\
&(V)&&\psi\to\lambda\psi\,,\quad\alpha_\text{MW}\to\lambda^{-1}\alpha_\text{MW} \,,
\end{aligned}
\end{equation}
where in each case $\lambda$ is some real positive number. As in \cite{Ammon:2009xh} we use (I) and (II) to set the boundary values of both $\sigma$ and $f$ to one, so that the metric will be asymptotically $AdS$. Also we can use (III) to set $r_h$ to one, but we will keep it as a bookkeeping device. We use (IV) to set the AdS radius $R$ to one. The relation (V) allows us to set $\alpha_\text{MW}=1$ by rescaling the baryon chemical potential, \ie we can relate states with different baryon chemical potentials in different theories characterized  by $\alpha_{\text{MW}}$ to each other.

A known solution of the equations of motion is the AdS Reissner-Nordstr\"om black hole,
\begin{equation}
\label{eq:RNsol}
\begin{aligned}
&\phi(r)=\mu_I-\frac{q_I}{r^2}\,,\quad\psi=\mu_B-\frac{q_B}{r^2}\quad\text{with}\quad q_i=\mu_i r_h^2\,,\\
&w(r)=0\,,\quad\sigma(r)=f(r)=1\,,\\
&N(r)=r^2-\frac{2m_0}{r^2}+\frac{2(\alpha_\text{YM}^2q_I^2+\alpha_\text{MW}^2q_B^2)}{3r^4}\quad\text{with}\quad m_0=\frac{r_h^4}{2}+\frac{\alpha_\text{YM}^2q_I^2+\alpha_\text{MW}^2q_B^2}{3r_h^2}\,.
\end{aligned}
\end{equation}

In order to obtain the solutions in the superfluid phase, \ie $w(r)\not\equiv0$, we have to resort to numerics. We will solve the equations of motion using a shooting method. We will vary the values of functions near the horizon until we find solutions with suitable values near the AdS boundary. We thus need the asymptotic forms of the solutions near the horizon $r=r_h$ and near the boundary $r\to\infty$.

Near the horizon, we expand all fields in powers of $\epsilon_h=r/r_h-1\ll 1$ with some constant coefficients. Three of these coefficients can be fixed as follows: We determine $r_h$ by the condition $N(r_h)=0$ which gives $m(r_h)=r_h^4/2$. Additionally, the time components of the gauge fields must be zero to obtain well-defined one-forms (see for example \cite{Kobayashi:2006sb}), \ie $\phi(r_h)=0$ and $\psi(r_h)=0$. The equations of motion then impose relations among the other coefficients. A straightforward exercise shows that only five coefficients are independent,
\begin{equation}
\label{eq:indcoeffhorizon}
\left\{\sigma_0^h,f_0^h,w_0^h,\phi_1^h,\psi_1^h\right\}\,,
\end{equation}
where the subscript denotes the order of $\epsilon_h$. All other near-horizon coefficients are determined in terms of these five independent coefficients.

Near the boundary, we expand all fields in powers of $\epsilon_b=(r_h/r)^2\ll 1$ with some constant coefficients. Again the equations of motion impose relations on these coefficients. There are seven independent coefficients
\begin{equation}
\label{eq:indcoeffbdy}
\left\{m_0^b,\phi_0^b,\phi_1^b,\psi_0^b,\psi_1^b,w_1^b,f_2^b\right\}\,,
\end{equation}
where here the subscript denotes the power of $\epsilon_b$. All other near-boundary coefficients are determined in terms of these seven independent coefficients.
We used the scaling symmetries \eqref{eq:scaltrans} to set $\sigma_0^b=f_0^b=1$. Our solutions will also have $w_0^b=0$ since we do not want to source the operator  $J_x^1$ in the dual field theory, \ie the $U(1)_3$ symmetry will be spontaneously broken. In our shooting method we choose a value of $\phi_0^b=\mu_I$, the isospin chemical potential, and of $\psi_0^b=\mu_B$, the baryon chemical potential, and then vary the five independent near-horizon coefficients until we find a solution which produces the desired values at the boundary.

In the following it will be often convenient to work with dimensionless coefficients by scaling out factors of $r_h$. We thus define the dimensionless functions $\mt(r)=m(r)/r_h^4$, $\phit(r)=\phi(r)/r_h$, $\psit(r)=\psi(r)/r_h$ and $\wt(r)=w(r)/r_h$, while $f(r)$ and $\sigma(r)$ are already dimensionless.
%-------------------------------------------------------------------------------------------------------------------------------------------
\subsection{Thermodynamics}
\label{sec:Thermodynamics}
%-------------------------------------------------------------------------------------------------------------------------------------------
In this section we extract thermodynamic information from our solutions. The gravity solutions describe thermal equilibrium in the boundary field theory. In order to extract thermodynamic quantities from the gravity solutions we can use well-known methods of black hole thermodynamics. 

The temperature $T$ in the boundary field theory is identified with the Hawking temperature of the black hole. The Hawking temperature for our black hole solutions is given by
\begin{equation}
\label{eq:temp}
T=\frac{\kappa}{2\pi}=\frac{r_h\sigma^h_0}{\pi}\left(1-\frac{\alpha_\text{YM}^2\left(\phi_1^h\right)^2+\alpha_\text{MW}^2\left(\psi_1^h\right)^2}{12\left(\sigma_0^h\right)^2}\right)\,,
\end{equation}
where $\kappa=\sqrt{\del_\mu\xi\del^\mu\xi}$ is the surface gravity of the black hole, with $\xi$ being the norm of the timelike Killing vector, and in the second equality we write $T$ in terms of the near-horizon coefficients. In the following we will often convert from the black hole radius $r_h$ to the temperature $T$ by inverting the above equation.

The entropy $S$ of the boundary field theory is identified with the Bekenstein-Hawking entropy of the black hole. For our ansatz we obtain
\begin{equation}
\label{eq:entropy}
S=\frac{2\pi}{\kappa_5^2}A_h=\frac{2V\pi r_h^3}{\kappa_5^2}=\frac{2\pi^4VT^3}{\kappa_5^2\left(\sigma_0^h\right)^3}\left(1-\frac{\alpha_\text{YM}^2\left(\phi_1^h\right)^2+\alpha_\text{MW}^2\left(\psi_1^h\right)^2}{12\left(\sigma_0^h\right)^2}\right)^{-3}\,,
\end{equation}
where $A_h$ is the area of the horizon and $V$ the spatial volume of the Minkowski space. 

The general statement of gauge/string duality which relates the field theory partition function to the string theory partition function may be used to calculate the thermodynamic potential of the boundary field theory, \ie in our case the grand potential. In the gravity approximation, which we use in this paper, the grand potential $\Omega$ is given as the temperature $T$ times the on-shell bulk action in Euclidean signature. We thus analytically continue to Euclidean signature and compactify the time direction with period $1/T$. We denote the Euclidean action as $I$ and its on-shell value as $I_\text{on-shell}$. Since our solutions are always static, we can integrate out the time direction which produces an overall factor of $1/T$. In order to simplify the expressions, we define $\It=I/T$. From now on we refer to $\It$ as the action. $\It$ splits into three parts, a bulk term, a Gibbons-Hawking term and counterterms,
\begin{equation}
\label{eq:eclideanaction}
\It=\It_\text{bulk}+\It_{GH}+\It_\text{ct}\,.
\end{equation}
The counterterms are needed to cancel the divergences of the bulk action and Gibbons-Hawking term which appear on-shell. To regulate these divergencies we introduce a hypersurface at $r=r_\text{bdy}$ with some large but finite $r_\text{bdy}$. On the field theory side $r_\text{bdy}$ corresponds to an UV cutoff. Ultimately we will remove the cutoff by taking $r_\text{bdy}\to\infty$.  Using the equations of motion, we obtain $\It_\text{bulk}^\text{on-shell}$ for our ansatz
\begin{equation}
\label{eq:Ibulkonshell}
\It_\text{bulk}^\text{on-shell}=\frac{V}{\kappa_5^2}\frac{1}{2f^2}rN\sigma(r^2f^2)'\Bigg|_{r=r_\text{bdy}}\,.
\end{equation}
For our ansatz, the Euclidean Gibbons-Hawking term is
\begin{equation}
\label{eq:GHterm}
\It_\text{GH}^\text{on-shell}=-\frac{1}{\kappa_5^2}\int\!\dd^3x\sqrt{\gamma}\;\nabla_\mu n^\mu=-\frac{V}{\kappa_5^2}N\sigma r^3\left(\frac{N'}{2N}+\frac{\sigma'}{\sigma}+\frac{3}{r}\right)\Bigg|_{r=r_\text{bdy}}\,,
\end{equation}
where $\gamma$ is the induced metric on the $r=r_\text{bdy}$ hypersurface and $n_\mu \dd x^\mu =1/\sqrt{N(r)}\;\dd r$ is the outward-pointing normal vector. The only divergence in the bulk action and Gibbons-Hawking term comes from the infinite volume of the asymptotically AdS space, hence, for our ansatz, the only nontrivial counterterm is
\begin{equation}
\label{eq:counterterm}
\It_\text{ct}^\text{on-shell}=\frac{3}{\kappa_5^2}\int\!\dd^3 x\sqrt{\gamma}=\frac{3V}{\kappa_5^2}r^3\sqrt{N}\sigma\Bigg|_{r=r_\text{bdy}}\,.
\end{equation}
Finally the grand potential $\Omega$ is given by
\begin{equation}
\label{eq:grandpotential}
\Omega=\lim_{r_\text{bdy}\to\infty}\It_\text{on-shell}\,.
\end{equation}
The baryon chemical potential $\mu_B$ is simply the boundary value of $\cala_t(r)=\psi(r)$ while the isospin chemical potential $\mu_I$ is the boundary value of $A_t^3(r)=\phi(r)$. The baryon charge density $\langle\calj_t\rangle$ and isospin charge density $\langle J_t^3\rangle$ of the dual field theory may be extracted from the on-shell action $\It_\text{on-shell}$ by
\begin{equation}
\label{eq:chargedensity}
\begin{split}
\langle\calj_t\rangle&=\frac{1}{V}\lim_{r_\text{bdy}\to\infty}\frac{\delta\It_\text{on-shell}}{\delta \cala_t(r_\text{bdy})}=-\frac{2\pi^3\alpha_\text{MW}^2T^3}{\kappa_5^2\left(\sigma_0^h\right)^3}\left(1-\frac{\alpha_\text{YM}^2\left(\phi_1^h\right)^2+\alpha_\text{MW}^2\left(\psi_1^h\right)^2}{12\left(\sigma_0^h\right)^2}\right)^{-3}\psit_1^b\,,\\
\langle J_t^3\rangle&=\frac{1}{V}\lim_{r_\text{bdy}\to\infty}\frac{\delta\It_\text{on-shell}}{\delta A_t^3(r_\text{bdy})}=-\frac{2\pi^3\alpha_\text{YM}^2T^3}{\kappa_5^2\left(\sigma_0^h\right)^3}\left(1-\frac{\alpha_\text{YM}^2\left(\phi_1^h\right)^2+\alpha_\text{MW}^2\left(\psi_1^h\right)^2}{12\left(\sigma_0^h\right)^2}\right)^{-3}\phit_1^b\,.
\end{split}
\end{equation}
Similarly, the current density $\langle J_x^1\rangle$ is
\begin{equation}
\label{eq:currentdensity}
\langle J_x^1\rangle=\frac{1}{V}\lim_{r_\text{bdy}\to\infty}\frac{\delta\It_\text{on-shell}}{\delta A_x^1(r_\text{bdy})}=-\frac{2\pi^3\alpha_\text{YM}^2T^3}{\kappa_5^2\left(\sigma_0^h\right)^3}\left(1-\frac{\alpha_\text{YM}^2\left(\phi_1^h\right)^2+\alpha_\text{MW}^2\left(\psi_1^h\right)^2}{12\left(\sigma_0^h\right)^2}\right)^{-3}\wt_1^b\,.
\end{equation}
The expectation value of the energy-momentum-tensor of the dual field theory is \cite{Balasubramanian:1999re,deHaro:2000xn}
\begin{equation}
\label{eq:defenergymomentumCFT}
\langle T_{ij}\rangle=\lim_{r_\text{bdy}\to\infty}\frac{2}{\sqrt{\gamma}}\frac{\delta\It_\text{on-shell}}{\delta\gamma^{ij}}=\lim_{r_\text{bdy}\to\infty}\left[\frac{r^2}{\kappa_5^2}\left(-K_{ij}+{K^l}_l\gamma_{ij}-3\gamma_{ij}\right)\right]_{r=r_\text{bdy}}\,,
\end{equation}
where $i,j,l\in\{t,x,y,z\}$ and $K_{ij}=1/2\sqrt{N(r)}\del_r\gamma_{ij}$ is the extrinsic curvature. We find
\begin{equation}
\label{eq:energymomentumCFT}
\begin{split}
\langle T_{tt}\rangle&=\frac{3\pi^4VT^4}{\kappa_5^2\left(\sigma_0^h\right)^4}\left(1-\frac{\alpha_\text{YM}^2\left(\phi_1^h\right)^2+\alpha_\text{MW}^2\left(\psi_1^h\right)^2}{12\left(\sigma_0^h\right)^2}\right)^{-4}\tilde m_0^b\,,\\
\langle T_{xx}\rangle&=\frac{\pi^4VT^4}{\kappa_5^2\left(\sigma_0^h\right)^4}\left(1-\frac{\alpha_\text{YM}^2\left(\phi_1^h\right)^2+\alpha_\text{MW}^2\left(\psi_1^h\right)^2}{12\left(\sigma_0^h\right)^2}\right)^{-4}\left(\tilde m_0^b-8f_2^b\right)\,,\\
\langle T_{yy}\rangle=\langle T_{zz}\rangle&=\frac{\pi^4VT^4}{\kappa_5^2\left(\sigma_0^h\right)^4}\left(1-\frac{\alpha_\text{YM}^2\left(\phi_1^h\right)^2+\alpha_\text{MW}^2\left(\psi_1^h\right)^2}{12\left(\sigma_0^h\right)^2}\right)^{-4}\left(\tilde m_0^b+4f_2^b\right)\,.
\end{split}
\end{equation}
For $\psi\equiv 0$ we recover the results obtained in \cite{Ammon:2009xh}. Notice that the energy-momentum tensor is still diagonal such that the momentum is zero even in the superfluid phase where the current $\langle J_x^1\rangle$ is non-zero. This result is guaranteed by our ansatz for the gauge fields which implies a diagonal Yang-Mills energy-momentum tensor and a diagonal metric. 

For $\tilde m_0^b=1/2+(\alpha_\text{YM}^2\mut_I^2+\alpha_\text{MW}^2\mut_B^2)/3$, $\sigma_0^h=1$, $\phit_1^h=2\mut_I$, $\psit_1^h=2\mut_B$, $f_2^b=0$, $\phit_b^0=\mut_I$, and $\psi_b^0=\mut_B$ we recover the correct thermodynamics for the Reissner-Nordstr\"om black hole, which preserves the $SO(3)$ rotational symmetry. For instance, we find that $\langle T_{xx}\rangle=\langle T_{yy}\rangle=\langle T_{zz}\rangle$ and $\Omega=-\langle T_{yy}\rangle$. For solutions with non-zero $\langle J_x^1\rangle$ the $SO(3)$ symmetry is spontaneously broken to $SO(2)$ and we find $\langle T_{xx}\rangle\not=\langle T_{yy}\rangle=\langle T_{zz}\rangle$. However we also find $\Omega=-\langle T_{yy}\rangle$ by just using the equations of motion as in \cite{Ammon:2009xh}.

Since the energy-momentum tensor is traceless (in Lorentzian signature), the dual field theory is scale invariant and describes a conformal fluid. The only physical parameters in the dual field theory are thus the ratios $\mu_B/T$, $\mu_I/T$ and $\mu_B/\mu_I$. Since only two of them are independent from each other, we choose $\mu_I/T$ and $\mu_B/\mu_I$ to determine the physical state of the boundary field theory in what follows.

%-------------------------------------------------------------------------------------------------------------------------------------------
\subsection{Phase transition and phase diagram}
\label{sec:Phase-diagram}
%-------------------------------------------------------------------------------------------------------------------------------------------
We expect a phase transition from the normal phase to a superfluid phase with a non-zero condensate $\langle J_x^1\rangle$ as the baryon and isospin chemical potential are varied. From \cite{Ammon:2009xh} we know that this phase transition occurs at zero baryon chemical potential. In the following we study the phase transition also at non-zero baryon chemical potential. 

\begin{figure}
\centering
\subfigure[]{\includegraphics[width=0.45\textwidth]{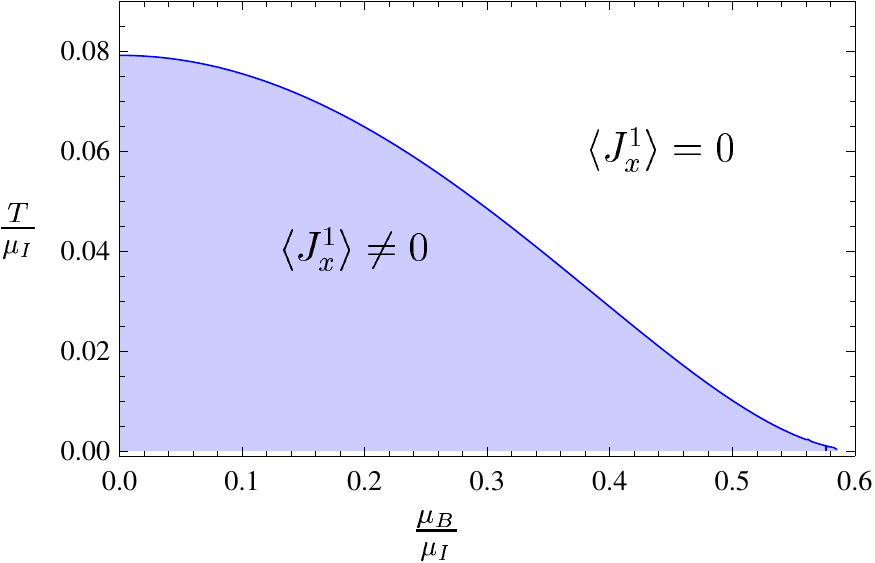}}
\hfill
\subfigure[]{\includegraphics[width=0.45\textwidth]{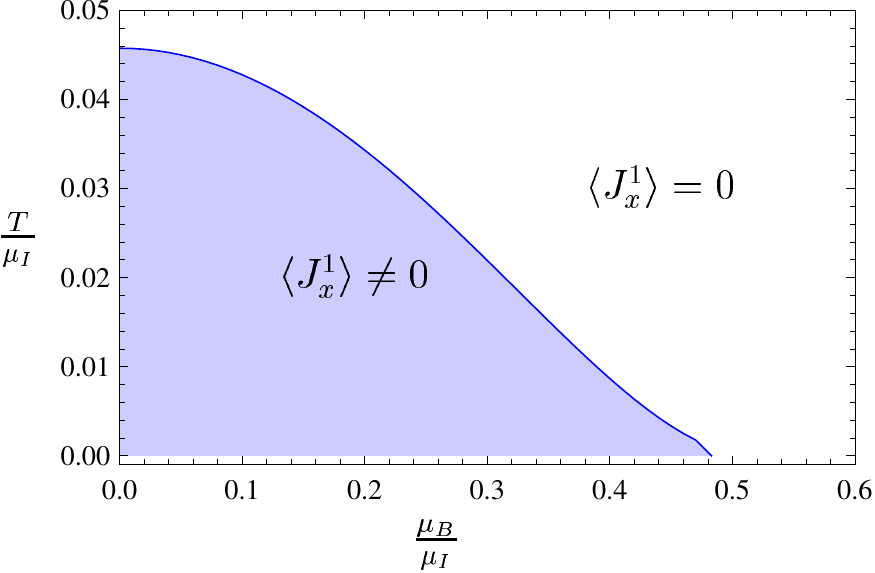}}
\caption{The phase diagram of the $U(2)$ Einstein Yang-Mills theory at finite temperature $T$, baryon $\mu_B$ and isospin chemical potential $\mu_I$ for $\alpha_{\text{YM}}=0.001$ (a) and $\alpha_{\text{YM}}=0.1$ (b): In the blue region the order parameter $\langle J_x^1\rangle$ is non-zero and the system is in the superfluid phase while in the white region the order parameter $\langle J_x^1\rangle$ is zero and the system is in the normal phase. }
\label{fig:phasediagramEYM}
\end{figure}

Let us first map out the phase diagram of the $U(2)$ EYM theory at finite temperature, baryon and isospin chemical potential for different values of the coupling constant $\alpha_{\text{YM}}$. We start our discussion for small $\alpha_{\text{YM}}$. Our numerical results are shown in fig.~\ref{fig:phasediagramEYM} and are confirmed by an analytic calculation at $\alpha_{\text{YM}}=0$ presented in section \ref{sec:The-semi-probe-limit}. In the blue region the order parameter $\langle J_x^1\rangle$ is non-zero and the system is in the superfluid phase while in the white region the order parameter $\langle J_x^1\rangle$ is zero and the system is in the normal phase. We observe that the phase boundary moves monotonically to lower temperatures compared to the isospin chemical potential $T/\mu_I$ as we increase the baryon chemical compared to the isospin chemical potential $\mu_B/\mu_I$. The order of the phase transition does not depend on the baryon chemical potential and stays second order. At a critical value for the ratio of baryon to isospin chemical potential $(\mu_B/\mu_I)_c$ we obtain a quantum critical point at zero temperature. In section \ref{sec:Zero-Temperature-Solutions-Backreact} we determine this critical ratio analytically. Its value can be found in \eqref{eq:QCP}. 

By increasing $\alpha_{\text{YM}}$ the area of the superfluid phase in the phase diagram decreases but the shape of the phase diagram stays the same until we reach a critical value for $\alpha_{\text{YM}}$. Beyond the critical value $(\alpha_{\text{YM}})_{c,1}=0.365\pm0.001$ we know from \cite{Ammon:2009xh} that the phase transition to the superfluid phase becomes first order at zero baryon chemical potential. If we now increase the baryon chemical potential, we find a critical point where the phase transition becomes second order again (for a sketch see figure~\ref{fig:sketchEYM} (b)). The phase transition at zero temperature is still continuous and therefore a quantum critical point. If we increase $\alpha_{\text{YM}}$, the critical point describing the change of the phase transition from first to second order moves to larger values of the ratio of baryon to isospin chemical potential. We find a critical value of $\alpha_{\text{YM}}$ where the zero temperature phase transition becomes first order and the quantum critical point disappears. Its value is given by $(\alpha_{\text{YM}})_{c,2}=0.492\pm0.008$. For $\alpha_{\text{YM}}$ above this value the phase transition is always first order (for a sketch see figure~\ref{fig:sketchEYM} (c)).

%%%%%%%%%%%%%%%%%%%%%%%%%%%%%%%% T H E     M O D E L  T  =  0
% !TEX root = SFisobar.tex

%-------------------------------------------------------------------------------------------------------------------------------------------
\subsection{Solutions at zero temperature}
\label{sec:Zero-Temperature-Solutions-Backreact}
%-------------------------------------------------------------------------------------------------------------------------------------------
In this section we consider the system exclusively at zero temperature. From the phase diagrams presented above we see that for large baryon compared to isospin chemical potential the system is in the normal state. Since the normal state is described by a Reissner-Nordstr\"om black hole, the zero temperature limit is an extremal Reissner-Nordstr\"om black hole. Zero temperature is given by fixing the isospin chemical potential,
\begin{equation}
\label{eq:zeroTempRS}
\mu_I = \frac{\sqrt{3 r_h^2 - \mu_B^2 \alpha_\text{MW}^2}}{\alpha_\text{YM}}\,.
\end{equation}
As usual this extremal black hole features an $AdS_2$ geometry in its near horizon region, \ie in the IR. The full solution in the near-horizon region is given by
\begin{equation}
\label{eq:nearhorizonRN}
\begin{split}
\dd s^2 & =-12 \xi^2\dd t^2+\frac{\dd \xi^2}{12 \xi^2}+r_h^2\dd \vec{x}^2\,,\\
\phi&=\frac{2\sqrt{3 r_h^2 - \mu_B^2 \alpha_\text{MW}^2}}{\alpha_\text{YM}r_h}\;\xi\,,\quad\psi=\frac{2\mu_B}{r_h}\xi\,,
\end{split}
\end{equation}
where $\xi=r-r_h$. According to the AdS/CFT dictionary, the dual field theory contains a one-dimensional CFT in the IR. Let us now consider this theory as we decrease the baryon chemical potential. From our numerical solutions we expect a phase transition towards a superfluid phase with non-zero vev $\langle J_x^1\rangle$. This phase transition should be triggered by an instability in the normal state. In order to obtain this instability we consider fluctuations of the gauge field $w(r)$ which is dual to the current $J_x^1$ about the extremal Reissner-Nordstr\"om background \cite{Iqbal:2010eh}. The equation of motion for this fluctuation is given by
\begin{equation}
\label{eq:eomwAdS2}
w''+\frac{2}{\xi}w'+\frac{3r_h^2-\mu_B^2\alpha_\text{MW}^2}{36\alpha_\text{YM}^2r_h^2\xi^2}w=0\,,
\end{equation}
where the prime denotes a derivative with respect to $\xi$. This equation is the equation of motion for a scalar field in $AdS_2$ with effective negative mass squared $m_\text{eff}^2=-\frac{3r_h^2-\mu_B^2\alpha_\text{MW}^2}{36\alpha_\text{YM}^2r_h^2}$. Thus according to the AdS/CFT dictionary, the IR dimension of the dual operator can be tuned by changing the baryon chemical potential. Hence, the fluctuation is stable until the mass is below the Breitenlohner-Freedman bound $m_\text{eff}^2=-1/4$.\footnote{Note that at the boundary the geometry is $AdS_5$ where the Breitenlohner-Freedman bound is $-4$.} In our case, the bound is given by
\begin{equation}
\label{eq:BFboundAdS2}
\frac{\sqrt{3r_h^2-\mu_B^2\alpha_\text{MW}^2}}{6\alpha_\text{YM}r_h}\le\frac{1}{2}\,.
\end{equation}
From this equation we may determine the baryon chemical potential at which the bound is saturated,
\begin{equation}
\label{eq:BFboundsatur}
\mu_B=\frac{r_h\sqrt{3-9\alpha_\text{YM}^2}}{\alpha_\text{MW}}\,.
\end{equation}
 With equation \eqref{eq:zeroTempRS}, we may determine the ratio between the baryon and the isospin chemical potential at which the Breitenlohner-Freedman bound is saturated. This ratio determines the point at which the system becomes unstable,
\begin{equation}
\label{eq:QCP}
\left(\frac{\mu_B}{\mu_I}\right)_c=\frac{\sqrt{1-3\alpha_\text{YM}^2}}{\sqrt{3}\;\alpha_\text{MW}}\,.
\end{equation}
Thus the Reissner-Nordstr\"om black hole may be unstable if $\alpha_\text{YM}<1/\sqrt{3}$ and a quantum critical point may exist if in addition $\alpha_\text{MW}$ is non-zero. This confirms our intuition obtained from our numerical results that at a given ratio of the baryon to isospin chemical potential a phase transition to a superfluid phase occurs. Unfortunately this calculation only determines the value for the ratio of baryon to isospin chemical potential where the system becomes unstable and not the phase boundary in general. For a continuous phase transition the two values coincide while for a first order phase transition, the transition always occurs before the instability is reached. Thus only for $\alpha_{\text{YM}}\le(\alpha_{\text{YM}})_{c,2}$ the phase boundary which is a quantum critical point and the critical value obtained here coincide. In \cite{Kaplan:2009kr} it is argued that the violation of the Breitenlohner-Freedman bound leads to a BKT-like transition.

Naively we may assume that the superfluid phase is non-degenerate at zero temperature and the entropy is zero. In the gravity dual this is translated to a zero horizon radius of the black hole. The solution with zero horizon radius differs from the zero temperature solutions described by the extremal limit of the AdS Reissner-Nordstr\"om black hole with finite horizon size. Similarly to \cite{Horowitz:2009ij,Basu:2009vv}, we choose the following ansatz which is consistent with the numerical results near $r=0$, namely
\begin{equation}
\label{eq:backreactZeroTAnsatz}
\begin{aligned}
&\phi \sim  \phi_1(r) \,,\quad &&\psi\sim  \psi_1(r) \,,\quad &&w \sim w_0 +\omega_1(r)\,,\quad &N \sim r^2 +N_1(r)\,,\\
&m \sim m_1(r)\,, &&\sigma\sim \sigma_0 +\sigma_1(r)\,, &&f \sim f_0 +f_1(r)\,,
\end{aligned}
\end{equation}
such that all fields with index one go to zero at $r=0$, e.g.  $f_0 +f_1(r) \rightarrow f_0$ as $r \rightarrow 0$. Plugging the ansatz above in \eqref{eq:eomsbr} and solving the equations of motion near the horizon $r=0$, we obtain the following solutions in the asymptotic forms
\begin{equation}
\label{eq:backreactZeroTSol}
\begin{aligned}
&\phi \sim  \phi_0 \sqrt{\frac{\beta}{r}}  e^{-\frac{\beta}{r}}\,,\quad &&N \sim r^2 - \frac{\alpha_{\text{YM}}^2 \beta^2\phi_0^2}{3\sigma_0^2}\frac{e^{-\frac{2\beta}{r}}}{r^2} \,,\quad &&w \sim w_0 \left(1-\frac{\phi_0^2}{4\sigma_0^2 \beta} \frac{e^{-\frac{2\beta}{r}}}{r} \right)\,,\\
&\psi = 0\,, &&\sigma \sim \sigma_0 \left(1+ \frac{\alpha_{\text{YM}}^2 \beta^2\phi_0^2}{6\sigma_0^2}\frac{e^{-\frac{2\beta}{r}}}{r^4}\right)\,, &&f \sim f_0 \left(1- \frac{\alpha_{\text{YM}}^2 \beta\phi_0^2}{12\sigma_0^2}\frac{e^{-\frac{2\beta}{r}}}{r^3}\right)\,,
\end{aligned}
\end{equation}
with $\beta = f_0^2 w_0$.  We can construct the full zero entropy solutions of the system by taking \eqref{eq:backreactZeroTSol}  as initial values near $r=0$ and integrate \eqref{eq:eomsbr} numerically  to the boundary using the shooting method. The result from that will describe the gravity dual of the superfluid ground state of the theory.  

It is important to note that a zero entropy solution is only consistent with $\psi$ being zero, \ie no baryon chemical potential. Thus the domain walls we can construct from this asymptotics always have zero baryon chemical potential and coincide with the one found in \cite{Basu:2009vv}. At finite baryon chemical potential we expect a solution which interpolates between the domain wall solutions at zero baryon chemical potential and the extremal Reissner-Nordstr\"om solution in the normal phase. This solution should always contain a black hole with finite horizon radius and thus its entropy increases with the baryon chemical potential. By dimensional analysis we obtain $S\propto V\mu_B^3$.

%%%%%%%%%%%%%%%%%%%%%%%%%%%%%%%% T H E     M O D E L   S E M I P R O B E   L I M I T
% !TEX root = SFisobar.tex

%-------------------------------------------------------------------------------------------------------------------------------------------
\subsection{The semi-probe limit}
\label{sec:The-semi-probe-limit}
%-------------------------------------------------------------------------------------------------------------------------------------------

\begin{figure}
 \centering
 \includegraphics[width=0.8\textwidth]{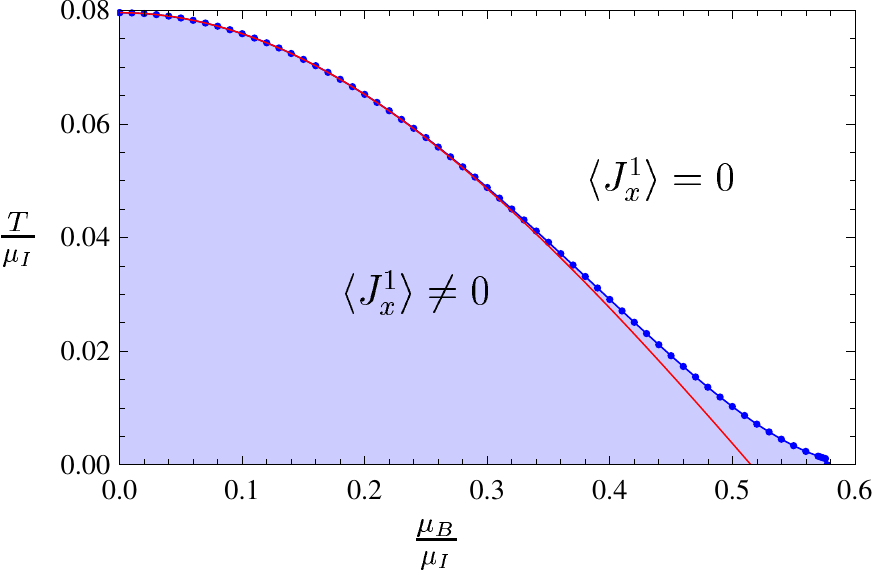}
 \caption{Phase diagram in the semi-probe limit: We compare the numerical data for the phase boundary (blue dots) with the analytic result (red line). We obtain a nice agreement for small baryon chemical potential where our approximation is valid. }
 \label{fig:phasediagramsemiprobe}
 \end{figure}

In this section we study the EYM system taking just the back-reaction of the $U(1)$ Maxwell field into account, \ie $\alpha_{\text{YM}}=0$. We call this limit the semi-probe limit. From equation \eqref{eq:QCP} we observe that there is the possibility of a quantum critical point at $\mu_B/\mu_I=1/\sqrt{3}\,\alpha_{\text{MW}}$ in this limit. In addition, the equations of motion \eqref{eq:eomsbr} simplify significantly and we can obtain an analytical solution if we restrict ourselves to small baryon chemical potential $\mu_B$ and small condensate $\langle J^1_x\rangle$. The equations of motion in the semi-probe limit read
\begin{equation}
\label{eq:eomsemiprobe}
\begin{aligned}
&m'=\frac{\alpha_{\text{MW}}^2r^3{\psi'}^2}{6}\,,\quad &&\psi''=-\frac{3}{r}\psi'\,,\\
&w''=-\frac{w\phi^2}{N^2}-w'\left(\frac{1}{r}+\frac{N'}{N}\right)\,,&&\phi''=\frac{w^2\phi}{r^2N}-\phi'\left(\frac{3}{r}\right)\,,
\end{aligned} 
\end{equation}
since $\sigma=f=1$ if the back-reaction of the $SU(2)$ Yang-Mills field is neglected. The equation for $m$ and $\psi$ can be integrated directly,
\begin{equation}
\label{eq:mpsisol}
\begin{split}
m&=\frac{r_h^4}{2}+\frac{\alpha_{\text{MW}}^2\mu_B^2 r_h^2}{3}\left(1-\frac{r_h^2}{r^2}\right)\,,\\
\psi&=\mu_B\left(1-\frac{r_h^2}{r^2}\right)\,.
\end{split}
\end{equation}
Thus we are left with the two equations of motion for the $SU(2)$ gauge fields in the given Reissner-Nordstr\"om background. By solving these equations numerically we can map out the phase diagram for $\alpha_{\text{YM}}=0$ (see fig.~\ref{fig:phasediagramsemiprobe}). The phase diagram looks similar to the one where a small back-reaction of the $SU(2)$ fields is included (see fig.~\ref{fig:phasediagramEYM}).

%-------------------------------------------------------------------------------------------------------------------------------------------
\subsubsection{The expansion}
\label{sec:The-Expansion}
%-------------------------------------------------------------------------------------------------------------------------------------------

In the limit of small $\mu_B$ and small $\langle J^1_x \rangle$, we can solve the equations of motion for $\phi$ and $w$ analytically. For the case $\mu_B=0$, this has already been done in \cite{Herzog:2009ci}. Similarly to \cite{Herzog:2010vz}, the solutions here are obtained as a double expansion in $\mu_B$ and $\langle J^1_x \rangle$ which are chosen to be proportional to the expansion parameters $\delta$ and $\epsilon$, respectively. More precisely, we choose $\delta \equiv \mut_B=\psit(\infty)$ and $\epsilon \equiv \wt_1^b \propto \langle J^1_x \rangle$  from \eqref{eq:currentdensity} where the tilde denotes dimensionless quantities which can be obtained by using \eqref{eq:scaltrans} to set $R=r_h=1$. We make the following ansatz for $\phit$ and $\wt$
\begin{equation}\label{ansatzphiw}
 \begin{split}
  \phit(r) =&~\phi_{0,0}(r)+\delta^2\,\phi_{2,0}(r)+\delta^4\,\phi_{4,0}(r)+\calo\left(\delta^6\right) \\
         & +\epsilon^2\left(\phi_{0,2}(r)+\delta^2\,\phi_{2,2}(r)\right)+\calo\left(\delta^4\epsilon^2\right) \\
         & +\epsilon^4\,\phi_{0,4}(r)+\calo\left(\delta^2\epsilon^4\right) \\
         	&+\calo\left(\epsilon^6\right) \,,\\
  \wt(r) =&~\epsilon\left(w_{0,1}(r)+\delta^2\,w_{2,1}(r)+\delta^4\,w_{4,1}(r)\right)+\calo\left(\delta^6\epsilon\right) \\
        & +\epsilon^3\left(w_{0,3}(r)+\delta^2\,w_{2,3}(r)\right)+\calo\left(\delta^4\epsilon^3\right) \\
        & +\epsilon^5 w_{0,5}(r)+\calo\left(\delta^2\epsilon^5\right) \,.
 \end{split}
\end{equation}
Inserting the ansatz \eqref{ansatzphiw} into \eqref{eq:eomsemiprobe}, we can construct a solution order by order in $\delta$ and $\epsilon$. The possible solutions are restricted by certain boundary conditions. At the horizon $r=1$, we demand that $\phit$ vanishes while $\wt$ has to be regular. At the boundary, $\wt$ is fixed to the expectation value $\langle J^1_x \rangle\propto \epsilon$ \eqref{eq:currentdensity} while the isospin chemical potential $\mut_{I}$ associated to $\phit$ receives finite corrections in $\delta$ and $\epsilon$. 

The coefficient functions to lowest order,  
\begin{equation}
 \begin{split}
\phi_{0,0}(r)=&~ 4\left(1-\frac{1}{r^2}\right) \, , \\
w_{0,1}(r)=&~ \frac{r^2}{(1+r^2)^2}
\end{split}
\end{equation}
are already known from \cite{Basu:2008bh} while the coefficients in the pure $\epsilon$ expansion, that is with $\delta=0$, were first computed in \cite{Herzog:2009ci}. To order $\epsilon^4$, they read 
\begin{equation}
 \begin{split}
 \phi_{0,2}(r)=&~\mu_{I0,2}\left(1-\frac{1}{r^2}\right)+\frac{5+7r^2-9r^4-3r^6}{96r^2\left(1+r^2\right)^3} \;,\quad \mu_{I0,2}=\frac{71}{6,\!720} \, , \\
    w_{0,3}(r)=&~ \frac{39 - 331r^2 -819 r^4 - 369r^6 + 156r^2(1+r^2)^3\,\text{ln}(1+\frac{1}{r^2})}{20,\!160(1+r^2)^5} \, , \\
 \phi_{0,4}(r)=&~\mu_{I0,4}\left(1-\frac{1}{r^2}\right) + \Phi_{0,4}(r)  \;,\quad \mu_{I0,4}= \frac{13\left(-4,\!015,\!679 + 5,\!147,\!520\,\text{ln}(2)\right)}{75,\!866,\!112,\!000} \, ,
 \end{split}
\end{equation}
where $\Phi_{0,4}(r)$ is a complicated function of $r$ which we do not write down explicitly here. The $ \mu_{Im,n}$ are determined by the regular boundary condition of $\wt$ at the horizon $r=1$ and describe corrections to the critical isospin chemical potential $\mut_I=4$ at $\delta^m \epsilon^n$ orders.
The lowest order coefficient functions in the pure $\delta$ expansion read
\begin{equation}
 \begin{split}
  \phi_{2,0}(r)=&~\mu_{I2,0}\left(1-\frac{1}{r^2}\right) \;,\quad  \mu_{I2,0}= \frac49\,\alpha_{\text{MW}}^2\left(-17+24\,\text{ln}(2)\right) \, ,\\
\phi_{4,0}(r)=&~\mu_{I4,0}\left(1-\frac{1}{r^2}\right)  \, ,\\
\mu_{I4,0}=&~ \frac{2}{243}\,\alpha_{\text{MW}}^2\big[-5,\!495 + 864\,\pi^2\,\text{ln}(2) + 192\,\text{ln}(2)\left(61 + 12\,\text{ln}(2)^2 -\text{ln}(8)\right) \\
            & - 13,\!824\,\text{Li}_3(1-i) - 13,\!824\,\text{Li}_3(1+i) + 12\,\zeta(3)\big] \, .
 \end{split}
\end{equation}
For small baryon chemical potential $\mut_B=\delta$, the critical isospin chemical potential for the phase transition will be corrected as
\begin{equation}
\label{eq:muIcorrected}
\mut_I^c(\delta)=4+\mu_{I2,0}\delta^2+\mu_{I4,0}\delta^4+{\cal O}(\delta^6) \, .
\end{equation}
This determines the phase boundary between the superfluid and the normal phase. We compare this analytic result with our numerical results in fig.~\ref{fig:phasediagramsemiprobe}. 

The lowest order coefficient functions in mixed orders read
\begin{equation}
 \begin{split}
\phi_{2,2}(r)=&~\mu_{I2,2}\left(1-\frac{1}{r^2}\right) + \Phi_{2,2}(r) \, , \\
\mu_{I2,2}=&~ \frac{\left(680,\!573 + 29,\!820\,\pi^2 - 404,\!232\,\text{ln}(2) - 1,\!406,\!160\,\text{ln}(2)^2\right)\alpha_{\text{MW}}^2}{6,\!350,\!400} \, ,
\\
 w_{2,1}(r)=&~ \alpha_{\text{MW}}^2\left(\frac{13 + r^2\left(7 + 6r^2 - 4\,\pi^2(1+r^2) + 24(1+r^2)\,\text{ln}(2)^2\right)}{9(1+r^2)^3}  \right.\\
           & \left. +\frac{4\left(3-20r^2 + 3r^4\right)\,\text{ln}(r)}{9(1+r^2)^2}- \frac{2\left(3 + 3r^4 + 4r^2\left(-5 + \text{ln}(64)\right)\right)\text{ln}(1+r^2)}{9(1+r^2)^2} \right. \\
          & \left. - \frac{16r^2\,\text{ln}(r)^2}{3(1+r^2)^2} - \frac{8r^2\left(\text{Li}_2(-r^2) + \text{Li}_2(1-r^2) - 2\,\text{Li}_2(\frac12(1-r^2)\right)}{3(1+r^2)^2} \right) \,,\\
 \end{split}
\end{equation}
where $\Phi_{2,2}(r)$ is a complicated function of $r$ which we do not display explicitly here.

\subsubsection{The free energy}

Using the results from the last section, we compute the contribution to the free energy up to order $\delta^m \epsilon^n$ for $m+n \leq 4$ from the gauge field term in the on-shell action 
\begin{equation}
\begin{split}
S &=-\frac{1}{4g_{\text{YM}}^2} \int\! \dd^5 x \sqrt{-g}\, F_{AB}^a F^{a AB} \\
&= \frac{\beta \,\mbox{Vol}_3 r_h^4}{2 g_{\text{YM}}^2}\int_1^{\infty}\!\dd r \left( r^3(\partial_r \phit)^2 - r \Nt(r) (\partial_r \wt)^2 + \frac{r}{\Nt(r)}(\phit \wt)^2\right) \\ 
&= \frac{\beta  \,\mbox{Vol}_3 r_h^4}{2 g_{\text{YM}}^2} \Big( r^3 \phit \left( \partial _r \phit \right)|_{r \rightarrow \infty} - \int_1^{\infty} \!\dd r\; r \Nt(r) \left(\partial_r \wt \right)^2 \Big) \, ,
\end{split}
\end{equation}
where $\mbox{Vol}_3$ is the spatial volume of the field theory and $\beta = 1/T$ is the inverse temperature. In the $\alpha_{\text{YM}} \rightarrow 0$ limit, only $\psi(r)$ contributes to the back-reaction which is described by 
\begin{equation}
\label{NSemi-probe}
\Nt(r)= r^2-\frac{1}{r^2}-\frac{2\left(r^2-1\right)}{3 r^4} \alpha_{\text{MW}}^2 \delta^2 \, ,
\end{equation}
where the expansion parameter $\delta \equiv \mut_B$  is chosen to be small.

For the background with vanishing condensate, i.e. $\omega(r)=0$, and 
\begin{equation}
\phit(r) = \frac{r^2-1}{r^2}\left(4+  \mu_{I0,2} \epsilon^2 + \mu_{I0,4} \epsilon^4 +\mu_{I2,0} \delta^2 +\mu_{I4,0} \delta^4 +\mu_{I2,2} \delta^2 \epsilon^2  \right)\, ,
\end{equation}
the on-shell action is
\begin{equation}
\begin{split}
S_{\rm vac} = \frac{\beta\, \mbox{Vol}_3 r_h^4}{ g_{\text{YM}}^2}  \Big[ 16 &+ 8 \mu_{I0,2} \epsilon^2 + \left( \mu_{I0,2}^2 +8 \mu_{I0,4} \right) \epsilon^4 +  8 \mu_{I2,0} \delta^2 + \left(\mu_{I2,0}^2+ 8 \mu_{I4,0} \right) \delta^4 \\
&+ 2\left(  \mu_{I0,2}\mu_{I2,0}+4\mu_{I2,2}\right) \delta^2 \epsilon^2 + {\cal O}\left(\delta^p \epsilon^q \right)\Big]\, ,
\end{split}
\label{S_onshell_vac}
\end{equation}
for $p+q = 6$.

For the background where $w \neq 0$ has condensed, the on-shell action reads
\begin{equation}
\label{S_onshell_sf}
\begin{split}
S_{\rm sf}  =  \frac{\beta\, \mbox{Vol}_3 r_h^4}{ g_{\text{YM}}^2}  \Big[ 16 &+ 8 \mu_{I0,2} \epsilon^2 + \left( \mu_{I0,2}^2 +8 \mu_{I0,4}+\frac{71}{215,\!040} \right) \epsilon^4 +  8 \mu_{I2,0} \delta^2 \\
&+ \left(\mu_{I2,0}^2+ 8 \mu_{I4,0} \right) \delta^4+ 2\left(  \mu_{I0,2}\mu_{I2,0}+4\mu_{I2,2}\right) \delta^2 \epsilon^2 + {\cal O}\left(\delta^p \epsilon^q \right) \Big]\, .
\end{split}
\end{equation}
\noindent
The difference in the values of the two on-shell actions is
\begin{equation}
\beta \Delta P = S_{\rm vac} - S_{\rm sf} = \frac{\beta \mbox{Vol}_3 r_h^4}{4 g_{\text{YM}}^2} \left(-\frac{71}{53,\!760} \epsilon^4 + {\mathcal O}\left(\delta^p \epsilon^q \right)  \right)\,.
\end{equation}
This result is known from \cite{Herzog:2009ci} which remains robust in our back-reacted background. 
The free energy in the grand canonical ensemble is minus the value of the on-shell action times the temperature, hence the quantity $\Delta P$ determines the difference in the free energy between the normal and superfluid phase. In this case, the free energy of the superfluid state is the smaller one because $\Delta P < 0$, and this implies the stability of the superfluid phase.

For small $\epsilon$ and small $\delta$, we have $\epsilon^4 \sim \left(\mut_I-\mut_I^c(\delta) \right)^2$. Using (III) in \eqref{eq:scaltrans} to restore dimensions by taking $\lambda = r_h \sim T$ \eqref{eq:temp}, the dimensionless $\mut_I$ will be replaced by $\frac{\mu_I}{r_h} \sim \frac{\mu_I}{T}$ and thus we have $\epsilon^2 \sim \left(T_c\left(\delta \right) - T \right)$.  The cancelation of the term proportional to $\delta^2\epsilon^2$ in the free energy difference suggests that the phase transition stays second order with mean field exponents as we increase the baryon chemical potential which coincides with our numerical result.

%%%%%%%%%%%%%%%%%%%%%%%%%%%%%%%% S E T U P          D 3 / D 7
% !TEX root = SFisobar.tex
%%%%%%%%%%%%%%%%%%%%%%%%%%%%%%%%%%%%%%%%%%%%%%%%%%%%%%%%%%%%%%%%%%%%%%%%%%%%
\section{D$3$/D$7$ Brane Setup}
\label{sec:d3/d7-brane-setup}
%%%%%%%%%%%%%%%%%%%%%%%%%%%%%%%%%%%%%%%%%%%%%%%%%%%%%%%%%%%%%%%%%%%%%%%%%%%%

%%%%%%%%%%%%%%%%%%%%%%%%%%%%%%%%%%%%%%%%%%%%%%%%%%%%%%%%%%%%%%%%%%%%%%%%%%%%
\subsection{Background and brane configuration} 
\label{sec:backgr-brane-conf}
%%%%%%%%%%%%%%%%%%%%%%%%%%%%%%%%%%%%%%%%%%%%%%%%%%%%%%%%%%%%%%%%%%%%%%%%%%%%
 In this section we investigate a string theory realization of the model studied above.   
We consider asymptotically $AdS_5\times S^5$ spacetime which is the near-horizon geometry of a stack of D$3$-branes. The $AdS_5\times
S^5$ geometry is holographically dual to the $\caln=4$ Super Yang-Mills
theory with gauge group $SU(N_c)$. The dual description of a finite
temperature field theory is an AdS black hole. We use the coordinates of
\cite{Kobayashi:2006sb} to write the AdS black hole background in Minkowski
signature as 
\begin{equation}
  \label{eq:AdSmetric}
  \dd s^2=\frac{\vrho^2}{2R^2}\left(-\frac{f^2}{\ft}\dd
    t^2+\ft
    \dd\vec{x}^2\right)+\left(\frac{R}{\vrho}\right)^2\left(\dd\vrho^2+\vrho^2\dd\Omega_5^2\right)\,,
\end{equation}
with $\dd\Omega_5^2$ the metric of the unit 5-sphere and
\begin{equation}
  f(\vrho)=1-\frac{\vrho_h^4}{\vrho^4},\quad
  \ft(\vrho)=1+\frac{\vrho_h^4}{\vrho^4}\,, 
\end{equation}
where $R$ is the AdS radius, with
\begin{equation}
  R^4=4\pi g_s N_c\,{\alpha'}^2 = 2\lambda\,{\alpha'}^2\,.
\end{equation}
The temperature of the black hole given by \eqref{eq:AdSmetric} may be
determined by demanding regularity of the Euclidean section. It is given by
\begin{equation}
  T=\frac{\vrho_h}{\pi R^2}\,. 
\end{equation}
In the following we may use the dimensionless coordinate
$\rho=\vrho/\vrho_h$, which covers the range from the event horizon at
$\rho=1$ to the boundary of the AdS space at $\rho\to\infty$.

To include fundamental matter, we embed $N_f$ coinciding D$7$-branes into
the ten-dimensional spacetime. These D$7$-branes host flavor gauge fields
$A_\mu$ with gauge group $U(N_f)$. This gauge field plays the same role as the gauge field in the Einstein-Yang-Mills systems. To write down the DBI action for the
D$7$-branes, we introduce spherical coordinates $\{r,\Omega_3\}$ in the
4567-directions and polar coordinates $\{L,\phi\}$ in the 89-directions
\cite{Kobayashi:2006sb}. The angle between these two spaces is denoted by
$\theta$ ($0\le\theta\le\pi/2$). The six-dimensional space in the
$456789$-directions is given by
\begin{equation}
\label{eq:coordiantesd3d7}
    \dd\varrho^2+\varrho^2\dd\Omega_5^2=\,\dd r^2+r^2\dd\Omega_3^2+\dd L^2+L^2\dd\phi^2=\,\dd\varrho^2+\varrho^2\left(\dd\theta^2+\cos^2\theta\dd\phi^2+\sin^2\theta\dd\Omega_3^2\right)\,,
\end{equation}
where $r=\varrho\sin\theta$, $\varrho^2=r^2+L^2$ and $L=\varrho\cos\theta$.
Due to the $SO(4)$ rotational symmetry in the 4567 directions, the embedding of the D$7$-branes only depends on the
radial coordinate $\rho$. Defining $\chi=\cos\theta$, we parametrize the
embedding by $\chi=\chi(\rho)$ and choose $\phi=0$ using the $SO(2)$
symmetry in the 89-direction. The induced metric $G$ on the D$7$-brane
probes is then
\begin{equation}
  \label{eq:inducedmetric}
    \dd s^2(G)=\frac{\vrho^2}{2R^2}\left(-\frac{f^2}{\ft}\dd
      t^2+\ft\dd\vec{x}^2\right)+\frac{R^2}{\vrho^2}\frac{1-\chi^2+\vrho^2(\del_\vrho\chi)^2}{1-\chi^2}\dd\vrho^2+R^2(1-\chi^2)\dd\Omega_3^2\,.
\end{equation}
The square root of the determinant of $G$ is given by
\begin{equation}
  \sqrt{-G}=\frac{\sqrt{h_3}}{4}\varrho^3f\ft(1-\chi^2)\sqrt{1-\chi^2+\varrho^2(\del_\varrho\chi)^2}\,,
\end{equation}
where $h_3$ is the determinant of the 3-sphere metric.

As in \cite{Erdmenger:2008yj} we split the $U(2)$ gauge symmetry on the
D$7$-brane into $U(1)_B\times SU(2)_I$ where the $U(1)_B$ describes the baryon charges and
$SU(2)_I$ isospin charges. As before we may introduce an isospin chemical potential $\mu_I$ as
well as a baryon chemical potential $\mu_B$ by introducing non-vanishing time component of the non-Abelian background fields.  Here we choose the generators of the $SU(2)_I$ gauge group to be the Pauli matrices $\sigma^i$ and the generator of the $U(1)_B$ gauge group to be $\sigma^0=\eins_{2\times 2}$. This non-zero time-components of the gauge fields $A_t^0=\cala_t$ and $A_t^3$ break the $U(2)$ gauge symmetry down to $U(1)_3$ generated by the third Pauli matrix $\sigma^3$. In order to study the transition to the superfluid state we additionally allow the gauge field $A_x^1$ to be non-zero. To obtain an isotropic and time-independent configuration in the field theory, the gauge field $A_x^1$ only depends on $\rho$. This leads to a similar ansatz for the gauge field as in the Einstein-Yang-Mills theory,
\begin{equation}
\label{eq:ansatzAD3D7}
A=\left(\cala_t(\vrho)\sigma^0+A_t^3(\vrho)\sigma^3\right)\dd t+A_x^1(\vrho)\sigma^1\dd x\,.
\end{equation}
With this ansatz, the field strength tensor on the branes has the following non-zero components,
\begin{equation}
  \label{eq:nonzeroF}
  \begin{split}
    &F^1_{\vrho x}=-F^1_{x\vrho}=\del_\vrho A^1_x\,,\\
    &F^2_{tx}=-F^2_{xt}=\frac{\gamma}{\sqrt{\lambda}}A^3_tA^1_x\,,\\
    &F^3_{\vrho t}=-F^3_{t\vrho}=\del_\vrho A^3_t\,,\\
    &F^0_{\vrho t}=-F^0_{t\vrho}=\del_\vrho \cala_t=\calf_{\vrho t}=-\calf_{t\vrho}\,.
  \end{split}
\end{equation}

%%%%%%%%%%%%%%%%%%%%%%%%%%%%%%%%%%%%%%%%%%%%%%%%%%%%%%%%%%%%%%%%%%%%%%%%%%%%%%%%%%%%%
\subsection{DBI action and equations of motion}
\label{sec:dbi-action-equations}
%%%%%%%%%%%%%%%%%%%%%%%%%%%%%%%%%%%%%%%%%%%%%%%%%%%%%%%%%%%%%%%%%%%%%%%%%%%%%%%%%%%%%
In this section we calculate the equations of motion which determine the
profile of the D$7$-brane probes and of the gauge fields on these branes.
The DBI action determines the shape of the brane embeddings, \ie the scalar
fields $\phi$, as well as the configuration of the gauge fields $A$ on these
branes. We consider the case of $N_f=2$ coincident D$7$-branes for which the
non-Abelian DBI action reads \cite{Myers:1999ps}
\begin{equation}
 \label{eq:non-AbelianDBI}
 S_{\text{DBI}}=-T_{D7}\:\str\int\!\dd^8\xi\:\sqrt{\det
     Q}\Bigg[\det\Big(P_{ab}\big[E_{\mu\nu}+E_{\mu i}(Q^{-1}-\delta)^{ij}E_{j\nu}\big]+2\pi\alpha'F_{ab}\Big)\Bigg]^{\frac{1}{2}}
\end{equation}
with
\begin{equation}
 \label{eq:defQDBI}
 Q^i{}_j=\delta^i{}_j+\ii 2\pi\alpha'[\Phi^i,\Phi^k]E_{kj}
\end{equation}
and $P_{ab}$ the pullback to the D$p$-brane, where for a D$p$-brane in $d$ dimensions we have 
$\mu,\,\nu=0,\dots, (d-1)$, $a,\,b=0,\dots, p$, $i,\,j = (p+1),\dots, (d-1)$,
$E_{\mu\nu} = g_{\mu\nu} + B_{\mu\nu}$. In our case we set $p=7$, $d=10$,
$B\equiv 0$. As in \cite{Erdmenger:2008yj} we can simplify this action
significantly by using the spatial and gauge symmetries present in our setup.
The action becomes
\begin{equation}
  \label{eq:DBI}
  \begin{split}
    S_{\text{DBI}}=&-T_{D7}\int\!\dd^8\xi\:\str\sqrt{|\det(G+2\pi\alpha'F)|}\\
    =&-T_{D7}\int\!\dd^8\xi\:\sqrt{-G}\:\str\Bigg[1+G^{tt}G^{\vrho\vrho}\left(\left(F^3_{\vrho
        t}\right)^2\left(\sigma^3\right)^2+2F^3_{\vrho t}\calf_{\vrho t}\sigma^3\sigma^0+\left(\calf_{\vrho
        t}\right)^2\left(\sigma^0\right)^2\right)\\
  &+G^{xx}G^{\vrho\vrho}\left(F^1_{\vrho x}\right)^2\left(\sigma^1\right)^2+G^{tt}G^{xx}\left(F^2_{tx}\right)^2\left(\sigma^2\right)^2\Bigg]^{\frac{1}{2}}\,,
  \end{split}
\end{equation}
where in the second line the determinant is calculated. Due to the
symmetric trace, all commutators between the matrices $\sigma^i$ vanish. It is
known that the symmetrized trace prescription in the DBI action is only valid
up to fourth order in $\alpha'$
\cite{Tseytlin:1997csa,Hashimoto:1997gm}. However the corrections to the higher order
terms are suppressed by $N_f^{-1}$ \cite{Constable:1999ac} (see also
\cite{Myers:2008me}). In \cite{Ammon:2008fc,Ammon:2009fe} we used two different approaches to
evaluate a non-Abelian DBI action similar to \eqref{eq:DBI}. First, we modified the
symmetrized trace prescription by omitting the commutators of the generators
$\sigma^i$ and then setting $(\sigma^i)^2=\eins_{2\times 2}$. This prescription makes the calculation of
the full DBI action feasible. Second, we expanded the non-Abelian DBI
action to fourth order in the field strength $F$.  We obtained the same physical properties
for the two approaches. We expect that the adapted symmetrized trace prescription also captures the relevant physics in this case such that we exclusively use the adapted symmetrized trace prescription in this paper. Using this prescription, the action becomes
 \begin{equation}
   \label{eq:DBIchangedstr}
     S_{\text{DBI}}=-\frac{T_{D7}}{4}\!\int\!\dd^8\xi\;\vrho^3f\ft(1-\chi^2)\left(\Upsilon_1(\rho,\chi,\At)+\Upsilon_2(\rho,\chi,\At)\right)\,,
 \end{equation}
with
\begin{equation}
  \label{eq:Upsilon}
  \begin{split}
    \Upsilon_i(\rho,\chi,\At)=\Bigg[&1-\chi^2+\rho^2(\del_\rho\chi)^2-\frac{2\ft}{f^2}(1-\chi^2)\left(\del_\rho
      \Xt_i\right)^2+\frac{2}{\ft}(1-\chi^2)\left(\del_\rho \At^1_x\right)^2\\
    &-\frac{\gamma^2}{2\pi^2\rho^4f^2}(1-\chi^2+\rho^2(\del_\rho\chi)^2)
    \left((\Xt_1-\Xt_2)\At_x^1\right)^2\Bigg]^{\frac{1}{2}}\,,
  \end{split}
\end{equation}
where the dimensionless quantities $\rho=\vrho/\vrho_h$ and
$\At=(2\pi\alpha')A/\vrho_h$ are used. The fields $X_1=\cala_t+A_t^3$ and
$X_2=\cala_t-A_t^3$ are the gauge fields on the $i$-th brane. In
\cite{Erdmenger:2008yj} it is shown that the non-Abelian DBI action with
$A_x^1=0$ decouples into two Abelian DBI actions in terms of these new gauge
fields $X_i$. To obtain first order equations of motion for the gauge fields
which are easier to solve numerically, we perform a Legendre transformation.
Similarly to \cite{Kobayashi:2006sb,Erdmenger:2008yj} we calculate the
electric displacement $p_i$ and the magnetizing field $p_x^1$ which are
given by the conjugate momenta of the gauge fields $X_i$ and $A_x^1$,
\begin{equation}
  \label{eq:conmomenta}
  p_i=\frac{\delta S_{\text{DBI}}}{\delta(\del_\vrho X_i)}\,,\qquad
  p_x^1=\frac{\delta S_{\text{DBI}}}{\delta(\del_\vrho A_x^1)}\,.
\end{equation}
In contrast to \cite{Kobayashi:2006sb,Karch:2007br,Mateos:2007vc,Erdmenger:2008yj}, the
conjugate momenta are not constant any more but depend on the radial coordinate
$\vrho$ due to the non-Abelian term $A_t^3A_x^1$ in the DBI action as in
\cite{Ammon:2008fc,Ammon:2009fe}. For the dimensionless momenta $\pt_i$ and $\pt_x^1$ defined as  
\begin{equation}
  \label{eq:pt}
  \pt=\frac{p}{2\pi\alpha'T_{D7}\vrho_h^3}\,,
\end{equation}
we get
\begin{equation}
  \label{eq:conmomentadim}
  \pt_i=\frac{\rho^3\ft^2(1-\chi^2)^2\del_\rho\Xt_i}{2f\Upsilon_i(\rho,\chi,\At)}\,,\quad\pt_x^1=-\frac{\rho^3f(1-\chi^2)^2\del_\rho\At_x^1}{2}
  \left(\frac{1}{\Upsilon_1(\rho,\chi,\At)}+\frac{1}{\Upsilon_2(\rho,\chi,\At)}\right)\,.
\end{equation}
Finally, the Legendre-transformed action is given by
\begin{equation}
  \label{eq:DBIlegendre}
  \begin{split}
    \tilde S_{\text{DBI}}&=S_{\text{DBI}}-\int\!\dd^8\xi\:
    \Bigg[
      \left(\del_\vrho X_i\right)\frac{\delta S_{\text{DBI}}}{\delta
        \left(\del_\vrho X_i\right)}+\left(\del_\vrho A_x^1\right)\frac{\delta S_{\text{DBI}}}{\delta\left(\del_\vrho A_x^1\right)}
    \Bigg]\\
    &=-\frac{T_{D7}}{4}\int\!\dd^8\xi\:\vrho^3f\ft(1-\chi^2)\sqrt{1-\chi^2+\rho^2(\del_\rho\chi)^2}\;V(\rho,\chi,\At,\pt)\,,
  \end{split}
\end{equation}
with
\begin{equation}
  \label{eq:V}
  \begin{split}
  V(\rho,\chi,\At,\pt)=&
  \left(1-\frac{\gamma^2}{2\pi^2\rho^4f^2}\left((\Xt_1-\Xt_2)\At_x^1\right)^2\right)^{\frac{1}{2}}\\
  &\times\left[ \left(\sqrt{1+\frac{8(\pt_1)^2}{\rho^6\ft^3(1-\chi^2)^3}}+\sqrt{1+\frac{8(\pt_2)^2}{\rho^6\ft^3(1-\chi^2)^3}}\right)^2-\frac{8(\pt^1_x)^2}{\rho^6\ft f^2(1-\chi^2)^3}\right]^{\frac{1}{2}}\,.
    \end{split}
\end{equation}
This action agrees with the one for finite baryon and isospin chemical potential (see \cite{Erdmenger:2008yj}) after $\pt_x^1\to 0$ and with the one for the superconducting state at pure isospin chemical potential (see \cite{Ammon:2008fc}) after $\pt_1\to -\pt_2$ and $\pt_x^1\to N_f \pt_x^1$. The change in $\pt_x^1$ has to be done such that the definitions agree in both cases.

Then the first order equations of motion for the gauge fields and their conjugate momenta are
\begin{equation}
  \label{eq:eomgauge}
  \begin{split}
    \del_\rho\Xt_i&=\frac{2f\sqrt{1-\chi^2+\rho^2(\del_\rho\chi)^2}}{\rho^3\ft^2(1-\chi^2)^2}\pt_iW(\rho,\chi,\At,\pt)U_i(\rho,\chi,\At,\pt)\, , \\
    \del_\rho\At_x^1&=-\frac{2\sqrt{1-\chi^2+\rho^2(\del_\rho\chi)^2}}{\rho^3f(1-\chi^2)^2}\pt_x^1W(\rho,\chi,\At,\pt)\, , \\
    \del_\rho\pt_{1/2}&=\pm\frac{\ft(1-\chi^2)\sqrt{1-\chi^2+\rho^2(\del_\rho\chi)^2}\gamma^2}{8\pi^2\rho
      fW(\rho,\chi,\At,\pt)}\left(\At_x^1\right)^2(\Xt_1-\Xt_2)\, , \\
    \del_\rho\pt_x^1&=\frac{\ft(1-\chi^2)\sqrt{1-\chi^2+\rho^2(\del_\rho\chi)^2}\gamma^2}{8\pi^2\rho
      fW(\rho,\chi,\At,\pt)}\left(\Xt_1-\Xt_2\right)^2\At_x^1\,,
  \end{split}
\end{equation}
with
\begin{equation}
  \label{eq:W}
  \begin{split}
    U_i(\rho,\chi,\At,\pt)&=\frac{\sqrt{1+\frac{8(\pt_1)^2}{\rho^6\ft^3(1-\chi^2)^3}}+\sqrt{1+\frac{8(\pt_2)^2}{\rho^6\ft^3(1-\chi^2)^3}}}{\sqrt{1+\frac{8(\pt_i)^2}{\rho^6\ft^3(1-\chi^2)^3}}}\,,\\
   W(\rho,\chi,\At,\pt)&=\sqrt{\frac{1-\frac{\gamma^2}{2\pi^2\rho^4f^2}\left((\Xt_1-\Xt_2)\At_x^1\right)^2}
   {\left(\sqrt{1+\frac{8(\pt_1)^2}{\rho^6\ft^3(1-\chi^2)^3}}+\sqrt{1+\frac{8(\pt_2)^2}{\rho^6\ft^3(1-\chi^2)^3}}\right)^2-\frac{8(\pt_x^1)^2}{\rho^6\ft f^2(1-\chi^2)^3}}}\,.
  \end{split}
\end{equation}
For the embedding function $\chi$ we get the second order equation of motion

\begin{equation}
  \label{eq:eomchi}
  \begin{split}
    \del_\rho
    \left[\frac{\rho^5f\ft(1-\chi^2)(\del_\rho\chi)V}{\sqrt{1-\chi^2+\rho^2(\del_\rho\chi)^2}}\right]\!
    =\!&-\frac{\rho^3f\ft\chi}{\sqrt{1-\chi^2+\rho^2(\del_\rho\chi)^2}}\Bigg\{
    \left[3\left(1-\chi^2\right)+2\rho^2(\del_\rho\chi)^2\right]V\\
    &-\frac{24\left(1-\chi^2+\rho^2(\del_\rho\chi)^2\right)}{\rho^6\ft^3\left(1-\chi^2\right)^3}W\!\Bigg[(\pt_1)^2U_1+(\pt_2)^2U_2-\frac{\ft^2}{f^2}(\pt_x^1)^2\Bigg]\!\Bigg\}\,.
  \end{split}
\end{equation}

We solve the equations of motion numerically by integrating them from the horizon at $\rho=1$ to the boundary $\rho=\infty$. The initial conditions may be determined by the asymptotic expansion of the gravity fields near the horizon
\begin{equation}
\label{eq:asymhor}
\begin{aligned}
&\Xt_i= &\frac{b_i}{(1 - \chi_0^2)^{\frac{3}{2}}B_i}(\rho-1)^2 &&+\calo\left((\rho-1)^3\right)\,,\\
&\At_x^1=a  & &&+\calo\left((\rho-1)^3\right)\,,\\
&\pt_{1/2}=b_{1/2}&\pm\frac{\gamma^2 a^2}{32\pi^2}\left(\frac{b_1}{B_1}-\frac{b_2}{B_2}\right)\left(B_1+B_2\right)(\rho-1)^2&&+\calo\left((\rho-1)^3\right)\,,\\
&\pt_x^1= & &&+\calo\left((\rho-1)^3\right)\,,\\
&\chi=\chi_0&-\frac{3\chi_0}{4B_1B_2}(\rho-1)^2&&+\calo\left((\rho-1)^3\right)\,,
\end{aligned}
\end{equation}
with
\begin{equation}
\label{eq:def}
B_i=\sqrt{1+\frac{b_i^2}{(1-\chi_0^2)^3}}\,.
\end{equation}
The terms in the asymptotic expansions are arranged according to their order in $(\rho-1)$. There are four independent parameters $\{a, b_1,b_2,\chi_0\}$ which have to be determined. In order to obtain the field theory quantities we determine the asymptotic expansion of the gravity fields near the AdS boundary
\begin{equation}
\label{eq:asymbdy}
\begin{aligned}
&\Xt_i=\mut_i& &-\frac{\dt_i}{\rho^2}& &+\calo\left(\rho^{-4}\right)\,,\\
&\At_x^1= & &+\frac{\dt_x^1}{2\rho^2}& &+\calo\left(\rho^{-4}\right)\,,\\
&\pt_i=\dt_i & & & &+\calo\left(\rho^{-4}\right)\,,\\
&\pt_x^1=\dt_x^1 & &-\frac{\gamma^2\dt_x^1(\mut_1-\mut_2)^2}{8\pi^2\rho^2} & &+\calo\left(\rho^{-4}\right)\,,\\
&\chi=&\frac{m}{\rho}& &+\frac{c}{\rho^3}&+\calo\left(\rho^{-4}\right)\,.
\end{aligned}
\end{equation}
Note that the factor of two in $\At_x^1$ is consistent with the earlier definitions in \cite{Ammon:2009fe} since here we have a different definition of the conjugate momenta (factor $N_f$). In this asymptotic expansion we find seven independent parameters $\{\mut_i,\dt_i,\dt_x^1,m,c\}$. Using the transformation of the gauge field from $\Xt_i$ to $\tilde\cala_t$ and $\At_t^3$, the independent parameters of the gauge fields $X_i$, $\{\mut_i,\dt_i\}$, may be translated into parameters of the asymptotic expansion of $\tilde\cala_t$ and $\At_t^3$,
\begin{equation}
\label{eq:transformparameters}
\begin{aligned}
&\mut_B=\mut_t^0=\frac{1}{2}(\mut_1+\mut_2)\qquad &\dt_B=\dt_t^0=\dt_1+\dt_2\,,\\
&\mut_I=\mut_t^3=\frac{1}{2}(\mut_1-\mut_2)&\dt_I=\dt_t^3=\dt_1-\dt_2\,.
\end{aligned}
\end{equation}
These parameters may be translated into field theory quantities according to the AdS/CFT dictionary (for details see \cite{Kobayashi:2006sb}): $\mu_B$ is the baryon chemical potential, $\mu_I$ the isospin chemical potential, 
\begin{equation}
\label{eq:defmut}
\mut_B=\sqrt{\frac{2}{\lambda}}\frac{\mu_B}{T}\,,\quad \mut_I=\sqrt{\frac{2}{\lambda}}\frac{\mu_I}{T}\,,
\end{equation}
the parameters $\dt$ are related to the vev of the flavor current $J$ by
\begin{equation}
\label{eq:defdt}
\dt_B=\dt_t^0=\frac{2^{\frac{5}{2}}\langle \calj_t\rangle}{N_c\sqrt{\lambda}T^3}\,,\quad \dt_I=\dt_t^3=\frac{2^{\frac{5}{2}}\langle J_t^3\rangle}{N_c\sqrt{\lambda}T^3}\,,\quad \dt_x^1=\frac{2^{\frac{5}{2}}\langle J_x^1\rangle}{N_c\sqrt{\lambda}T^3}\,,
\end{equation}
and $m$ and $c$ to the bare quark mass $M_q$ and the quark condensate $\langle\bar\psi\psi\rangle$, 
\begin{equation}
\label{eq:defmc}
m=\frac{2M_q}{\sqrt{\lambda}T},\quad c=-\frac{8\langle\bar\psi\psi\rangle}{\sqrt{\lambda}N_fN_cT^3}\,,
\end{equation}
respectively.  There are three independent physical parameters, \eg $m$, $\mu_B$ and $\mu_I$ in the grand canonical ensemble. The asymptotic expansion close to the horizon has four independent solutions. These parameters may be fixed by choosing the three independent physical parameters, \ie the state in the field theory and by the constraint that $\At_x^1$ goes to zero at the boundary, \ie the $U(1)_3$ symmetry is spontaneously broken. We use a standard shooting method to determine the parameters at the horizon.

%%% Local Variables: 
%%% mode: latex
%%% TeX-master: "SFisobar"
%%% End: 

%%%%%%%%%%%%%%%%%%%%%%%%%%%%%%%% P H A S E T R A N S I T I O N
% !TEX root = SFisobar.tex
%-------------------------------------------------------------------------------------------------------------------------------------------
\subsection{Thermodynamics}
\label{sec:Thermodynamicsd3d7}
%-------------------------------------------------------------------------------------------------------------------------------------------

In this section we study the contribution of the D$7$-branes to the thermodynamics. According to the AdS/CFT dictionary the partition function $Z$ of the boundary field theory is given in terms of the Euclidean on-shell supergravity action $I_\text{on-shell}$,
\begin{equation}
\label{eq:partfunc}
	Z=\ee^{-I_\text{on-shell}}\,.
\end{equation}
Thus the thermodynamic potential, \ie the grand potential in the grand canonical ensemble, is proportional to the Euclidean on-shell action
\begin{equation}
\label{eq:defgrandpot}
	\Omega=-T\ln Z=T I_\text{on-shell}\,.
\end{equation}
To calculate the contribution of the D$7$-branes to the grand potential, we have to determine the Euclidean version of the DBI-action \eqref{eq:DBIchangedstr} on-shell. For this purpose, we first perform a Wick rotation in the time direction. Next we renormalize the action by adding appropriate counterterms $I_\text{ct}$ (see \cite{Skenderis:2002wp} for a review and \cite{Karch:2005ms} for probe D-branes). In our case the counterterms are the same as in \cite{Kobayashi:2006sb,Mateos:2007vn,Erdmenger:2008yj},
\begin{equation}
\label{eq:counterterms}
	I_\text{ct}=-\frac{\lambda N_c N_f V_3 T^3}{128}\left[\left(\rho^2_\text{max}-m^2 \right)^2-4mc \right]\,,
\end{equation}
where $\rho_\text{max}$ is the UV-cutoff and $V_3$ the Minkowski space volume. Then the renormalized Euclidean on-shell action $I_R$ may simply be written as
\begin{equation}
\label{eq:renomralizedaction}
I_R=\frac{\lambda N_c N_f V_3 T^3}{32}\left(\frac{G(m,\mut)}{N_f}-\frac{1}{4}\left[\left(\rho_\text{min}^2-m^2\right)^2-4mc \right] \right) \,,
\end{equation}
where $\rho_\text{min}$ determines the minimal value of the coordinate $\rho$ on the D$7$-branes, \ie $\rho_\text{min}=1$ for black hole embeddings which we consider exclusively in this paper and
\begin{equation}
\label{eq:defG}
G(m,\mut)=\int_{\rho_\text{min}}^\infty\!\dd\rho\left[\rho^3 f\ft(1-\chi^2)\left(\Upsilon_1(\rho,\chi,\At)+\Upsilon_2(\rho,\chi,\At)\right)-N_f\left(\rho^3-\rho m\right)\right]\,.
\end{equation}
In the following we consider the dimensionless grand potential $\calw_7$ defined as
\begin{equation}
\label{eq:defgrandpot7}
\Omega_7=TI_R=\frac{\lambda N_cN_f V_3 T^4}{32}\calw_7\,.
\end{equation}
By considering the variation of the grand potential with respect to the gravity fields, it can be shown (see \cite[section 5.3]{Ammon:2009fe} for the pure isospin case) that the above definition of the grand potential is consistent and that the order parameter $\dt_x^1$ is not a thermodynamical variable. 

%-------------------------------------------------------------------------------------------------------------------------------------------
\subsubsection{Phase transition and phase diagram}
\label{sec:Phase-diagramd3d7}
%-------------------------------------------------------------------------------------------------------------------------------------------
We expect that a phase transition occurs between a normal fluid phase and a superfluid phase. At zero baryon chemical potential we know from \cite{Ammon:2008fc,Ammon:2009fe} that the phase transition from the normal phase to the superfluid phase is second order with mean field exponents. In this section we consider the phase transition at non-zero baryon chemical potentials.

\begin{figure}[htbp]
\centering
\includegraphics[width=0.7\textwidth]{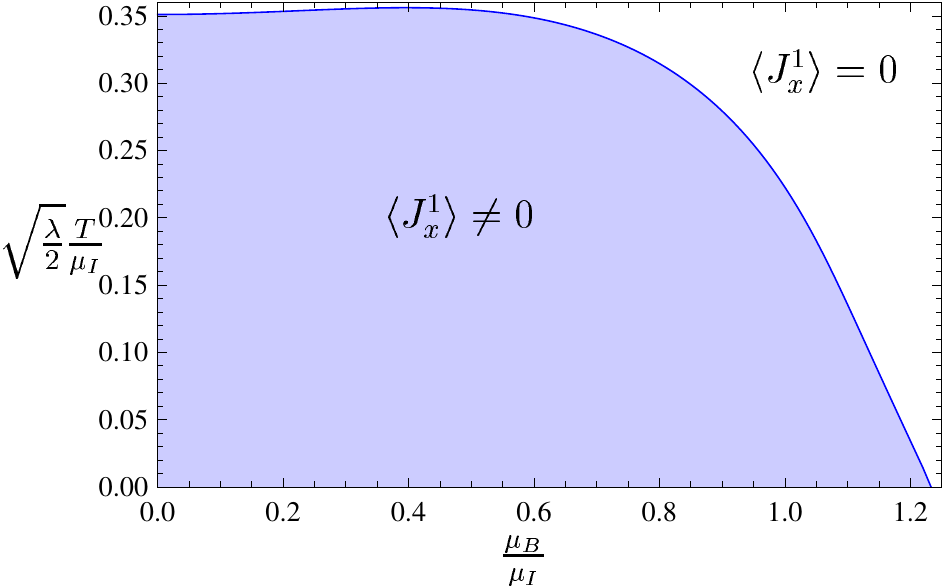}
\caption{The phase diagram for fundamental matter in thermal strongly-coupled $\caln = 2$ SYM theory at zero quark mass with $\mu_I$ the isospin chemical potential, $\mu_B$ the baryon chemical potential, $T$ the temperature and $\lambda$ the 't Hooft coupling: In the blue region the order parameter $\langle J_x^1\rangle$ is non-zero and the system is in the superfluid phase while in the white region the order parameter  $\langle J_x^1\rangle$ is zero and the system is in the normal phase.}
\label{fig:phasediad3d7}
\end{figure}

First we map out the phase diagram of the given theory with zero quark mass $m=0$.  The phase diagram is shown in fig.\,\ref{fig:phasediad3d7}.  As we increase the baryon chemical potential the transition temperature to the superfluid phase first increases. For $\mu_B/\mu_I\gtrsim0.4$ the transition temperature monotonically decreases to zero as the baryon chemical potential grows. We can show numerically that the phase transition is always second order. In the next subsection we will show numerically that the transition temperature is zero at $\mu_B/\mu_I\approx 1.23$, and hence we obtain a quantum critical point.

%%%%%%%%%%%%%%%%%%%%%%%%%%%%%%%% Z E R O    T E M P E R A T U R E
% !TEX root = SFisobar.tex

%-------------------------------------------------------------------------------------------------------------------------------------------
\subsection{Zero temperature solutions and quantum critical point}
\label{sec:Zero-Temperature-Solutions}
%-------------------------------------------------------------------------------------------------------------------------------------------
In this section we study the D$3$/D$7$-brane setup at zero temperature. The zero temperature limit is given by $\rho_h\to 0$, \ie $f=\ft=1$. The induced metric on the D$7$-branes may now be written in $(L,r)$ coordinates (see equation \eqref{eq:coordiantesd3d7})
\begin{equation}
\label{eq:inducedmetriczeroT}
\dd s^2=\frac{r^2+L^2}{2R^2}\left(-\dd t^2+\dd \vec{x}^2\right)+\frac{R^2}{r^2+L^2}\left(1+(\del_r L)^2\right)\dd r^2+\frac{R^2r^2}{r^2+L^2}\dd\Omega_3^2\,.
\end{equation}
The square root of the metric is now
\begin{equation}
\label{eq:sqrtmetriczeroT}
\sqrt{-G}=\frac{\sqrt{h_3}}{4}r^3\sqrt{1+(\del_r L)^2}\,,
\end{equation}
and using the adapted symmetrized trace prescription, the DBI action becomes
\begin{equation}
\label{eq:DBIzeroT}
S_{\text{DBI}}=-T_{D7}\int\!\dd\xi^8\frac{r^3}{4}\left[\Xi_1(\rt,\Lt,\At)+\Xi_2(\rt,\Lt,\At)\right]\,,
\end{equation}
with
\begin{equation}
\label{eq:Xi}
\begin{split}
\Xi_i(\rt,\Lt,\At)=\Bigg[&1+(\del_{\rt} \Lt)^2-2(\del_{\rt} \Xt_i)^2+2(\del_{\rt} \At_x^1)^2\\
	&-\frac{\gamma^2}{2\pi^2(\rt^2+\Lt^2)^2}(1+(\del_{\rt} \Lt)^2)\left((\Xt_1-\Xt_2)\At_x^1\right)^2\Bigg]^{\frac{1}{2}}\,,
\end{split}
\end{equation}
where the dimensionless quantities are now defined by
\begin{equation}
\label{eq:dimensionlessquantzeroT}
\rt = \frac{r}{R}, \quad \Lt=\frac{L}{R}, \quad \At = \frac{2 \pi \alpha'}{R} A \,.
\end{equation}
In the normal phase, \ie $A_x^1\equiv 0$, the equations of motion for the gauge fields $X_i$ for the massless embedding $L=0$,
\begin{equation}
\label{eq:eomsnormalphasezeroT}
\del_{\rt}\Xt_i=\frac{2\dt_i}{\sqrt{\rt^6+8\dt_i^2}}\,
\end{equation}
can be solved analytically \cite{Karch:2007br}. The solution expressed in terms of incomplete Beta functions is given by
\begin{equation}
\label{eq:solnormalphasezeroT}
\Xt_i(\rt)= \frac{\dt_i^\frac{1}{3}}{6} B\left(\frac{\rt^6}{8\dt_i^2+\rt^6};\frac{1}{6},\frac{1}{3} \right) .  
\end{equation}
From the asymptotic form near the boundary, we can read off the chemical potential and the density
(see equation \eqref{eq:asymbdy}),
\begin{equation}
\label{eq:asymnormalphasezeroT}
\Xt_i=\frac{\dt_i^\frac{1}{3}2\sqrt{\pi}}{\sqrt{3}}\frac{ \Gamma\left(\frac{7}{6}\right)}{\Gamma\left(\frac{2}{3}\right)} - \frac{\dt_i}{\rt^2}+\cdots\,.
\end{equation}
In the normal fluid phase we may now consider fluctuations $Z_\pm=A_x^1\pm \ii A_x^2$ (see \cite{Erdmenger:2007ja,Erdmenger:2008yj} for more details) and look for an instability which may lead to a phase transition. For the flat embedding $L=0$, the equation of motion for the fluctuation $Z_\pm$ at zero momentum is given by
\begin{eqnarray}
\label{eq:eomfluczeroT}
\frac{\dd^2\tilde{Z_+}(\rt)}{\dd\rt^2} +\frac{F'(\rt)}{F(\rt)} \frac{\dd \tilde{Z_+}(\rt)}{\dd\rt}+\frac{8}{\rt^4}\left(\tilde{\omega}+\frac{\gamma}{4\sqrt{2}\pi} \left(\tilde{X_1}-\tilde{X_2} \right)\right)^2 \tilde{Z_+}(\rt)&=&0 \,,\\
\frac{\dd^2\tilde{Z_-}(\rt)}{\dd\rt^2} +\frac{F'(\rt)}{F(\rt)} \frac{\dd \tilde{Z_-}(\rt)}{\dd\rt}+\frac{8}{\rt^4}\left(\tilde{\omega}-\frac{\gamma}{4\sqrt{2}\pi} \left(\tilde{X_1}-\tilde{X_2} \right)\right)^2 \tilde{Z_-}(\rt)&=&0\, ,
\end{eqnarray}
with $\tilde{\omega}=\sqrt{\frac{\lambda}{2}}~\alpha'\omega$ coming from the ansatz  $A_x^i(\rt,t) = A_x^i (\rt) e^{-i\omega t}$.
The analytical expression for $\tilde{X_i}(\rt)$ is given in \eqref{eq:solnormalphasezeroT} and 
\begin{equation}
F(\rt)= \rt^3 \left(\frac{1}{\sqrt{1-2\left(\partial_{\rt} \tilde{X}_1\right)^2}}  +\frac{1}{\sqrt{1-2\left(\partial_{\rt} \tilde{X}_2\right)^2}} \right). 
\end{equation}
The system is unstable if the imaginary part of the quasinormal frequency is positive. For a massless embedding $L\equiv 0$, we find this instability at $\mu_B/\mu_I\approx 1.23$. 

%-------------------------------------------------------------------------------------------------------------------------------------------
\subsubsection{What is the origin of the instability?}
\label{sec:What-is-the}
%-------------------------------------------------------------------------------------------------------------------------------------------

In the back-reacted Einstein-Yang-Mills theory we see that the instability of the extremal Reissner-Nordstr\"om black hole which triggers the phase transition to the superfluid phase can be related to the violation of the Breitenlohner-Freedman bound of the field $A_x^1$ in the near horizon geometry of the black hole. This near horizon geometry features an $AdS_2$ factor (see section \ref{sec:Zero-Temperature-Solutions-Backreact}). In this section we investigate the instability in the D$3$/D$7$ brane setup at zero temperature. In the following we show that in contrast to the back-reacted Einstein-Yang-Mills system, the instability in the brane setup cannot be related to the violation of the Breitenlohner-Freedman bound in the far IR, \ie the fluctuation of the field $A_x^1$ does not become unstable in the far IR. In this system we expect that the instability occurs in the bulk.

Let us now consider a fluctuation $A_x^1$ about the solution in the normal phase given by equation \eqref{eq:solnormalphasezeroT} for zero quark mass, \ie $L\equiv 0$. The equation of motion for this fluctuation is given by
\begin{equation}
\label{eq:eomflucAx1zeroT}
(\At_x^1)''+\frac{F'(\rt)}{F(\rt)}(A_x^1)'+\frac{\gamma^2(\Xt_1-\Xt_2)^2}{4\pi^2\rt^4}\At_x^1=0\,.
\end{equation}
As in section \ref{sec:Zero-Temperature-Solutions-Backreact} we consider the equation of motion in the far IR. Since there is no horizon at zero temperature in the brane setup, the expansion is around $\rt=0$. The equation of motion becomes trivial, $\del_{\rt}^2\At_x^1=0$, if we expand to first order only. Thus $\At_x^1/r$ satisfies the equation of motion of a massless scalar in $AdS_2$. In contrast to the EYM setup, the mass cannot be tuned by changing a UV quantity and the Breitenlohner-Freedman bound can never be violated\footnote{We thank K. Jensen for pointing this out.}. Hence the fluctuation is stable in the IR. We expect that the instability observed by the numerical study of the quasinormal modes above occurs along the flow. Thus the origin of the instability in this brane setup is different compared to the one in the back-reacted Einstein-Yang-Mills system studied in section \ref{sec:Zero-Temperature-Solutions-Backreact}. 

%%%%%%%%%%%%%%%%%%%%%%%% C O N C L U S I O N
% !TEX root = SFisobar.tex

%-------------------------------------------------------------------------------------------------------------------------------------------
\section{Conclusion}
\label{sec:Conclusion}
%-------------------------------------------------------------------------------------------------------------------------------------------
We have considered holographic models of field theories with global $U(2)$ symmetry. The $U(2)$ symmetry allows us to switch on two chemical potentials: a baryon and an isospin chemical potential. Holographically we realized the global $U(2)$ symmetry in two different ways: first by an $U(2)$ Einstein-Yang-Mills theory and second by the D$3$/D$7$ brane setup with two coincident D$7$-branes. We mapped out the phase diagrams for both systems and found interesting similarities and differences which we already discussed in the introduction.

It would be interesting to study the origin of these differences in the phase diagrams in more detail. For example a detailed analysis of how the order of the phase transition can be changed by varying the form of the interaction would be attractive. In addition the study of back-reaction effects in the D$3$/D$7$ model may lead to new behavior in the phase diagrams. Furthermore it is important to establish a full understanding of the instability mechanism in the D$3$/D$7$ brane setup and its difference to the violation of the Breitenlohner-Freedman bound found in the Einstein-Yang-Mills theory. This may lead to a characterization of the universality classes of quantum phase transitions.

%%%%%%%%%%%%%%%%%%%%%%%% A C K N O W L E D G M E N T S
\section*{Acknowledgements}
%\addcontentsline{toc}{section}{Acknowledgments}
We are grateful to M. Ammon, A. O'Bannon and D. Son for discussions and especially to K. Jensen for several comments on the manuscript. Moreover we thank Y.-I. Shin, C. H. Schunck, A. Schirotzek, and W. Ketterle, as well as L.-Y. He, M. Jin, and P.-F. Zhuang for the permission to reproduce the figures.  This work was supported in
part by  {\it The Cluster of Excellence for Fundamental Physics - Origin and
  Structure of the Universe}.

%%%%%%%%%%%%%%%%%%%%%%%% R E F E R E N C E S
\providecommand{\href}[2]{#2}\begingroup\raggedright\endgroup

%\bibliographystyle{felice_utcaps}
%\bibliography{library}

\end{document}